\documentstyle[prb,aps,epsf,twocolumn]{revtex}
\newcommand{\eq}{\begin{equation}}
\newcommand{\ee}{\end{equation}}
\newcommand{\s}{{\sigma}}

\newcommand{\Xb}{{\bar{X}}}

\newcommand{\va}{{\vec{a}}}
\newcommand{\vrr}{{\vec{r}}}

\newcommand{\vq}{{\vec{q}}}
\newcommand{\vs}{{\vec{s}}}
\newcommand{\vS}{{\vec{S}}}
\newcommand{\vQ}{{\vec{Q}}}

\newcommand{\p}{{\partial}}
\newcommand{\gr}{{\nabla}}
\newcommand{\ra}{{\rightarrow}}
\def\lam{\lambda}
\def\t{\theta}
\def\d{{\delta}}
\def\rhot{{\tilde{\rho}}}

\def\a{\alpha}

\def\e{\epsilon}
\def\o{{\omega}}
\def\sul{\sum\limits}
\def\dd{d^{\dagger}}

\def\half{{1\over2}}
\def\third{{1\over3}}
\def\twof{{2\over5}}
\def\threes{{3\over7}}

\def\rhob{{\bar \rho}}
\def\intq{{\int {d^2q\over(2\pi)^2}}}

\def\hX{{\hat X}}
\def\hH{{\hat H}}
\def\hV{{\hat V}}
\def\hS{{\hat S}}
\def\hY{{\hat Y}}
\def\ua{\uparrow}
\def\da{\downarrow}
\def\eqa{\begin{eqnarray}}
\def\eea{\end{eqnarray}}
\def\prl{{Phys. Rev. Lett.\ }}
\def\prb{{Phys. Rev. {\bf B}}}
\def\jpc{{Jour. Phys. {\bf C}}}

\begin{document}
\draft
\flushbottom
\twocolumn[
\hsize\textwidth\columnwidth\hsize\csname @twocolumnfalse\endcsname
\title{ Hamiltonian Description of Composite Fermions: Magnetoexciton Dispersions }
\author{  Ganpathy Murthy}
\address{
Physics Department, Boston University, Boston MA 02215\\
and
Department of Physics and Astronomy, Johns Hopkins University,
Baltimore MD 21218}
\date{\today}
\maketitle
\tightenlines
\widetext
\advance\leftskip by 57pt
\advance\rightskip by 57pt

\begin{abstract}
A microscopic Hamiltonian theory of the FQHE, developed by Shankar and
myself based on the fermionic Chern-Simons approach, has recently been
quite successful in calculating gaps in Fractional Quantum Hall
states, and in predicting approximate scaling relations between the
gaps of different fractions. I now apply this formalism towards
computing magnetoexciton dispersions (including spin-flip dispersions)
in the $\nu=\third$, $\twof$, and $\threes$ gapped fractions, and find
approximate agreement with numerical results.  I also analyse the
evolution of these dispersions with increasing sample thickness,
modelled by a potential soft at high momenta.  New results are
obtained for instabilities as a function of thickness for $\twof$ and
$\threes$, and it is shown that the spin-polarized $\twof$ state, in
contrast to the spin-polarized $\third$ state, cannot be described as
a simple quantum ferromagnet.
\end{abstract}
\vskip 1cm

]
\narrowtext
\tightenlines
\section{Introduction}
\label{sec1}

The Fractional Quantum Hall (FQH) effect\cite{fqhe-ex} has introduced
us to new states of electrons in high magnetic fields. Seminal
theoretical progress was made by Laughlin\cite{laugh}, who showed that
for filling fractions $\nu=1/(2p+1)$ the electrons form a strongly
correlated state which is incompressible, and has quasiparticle
excitations with fractional charge\cite{laugh}
$e^*=e/(2p+1)$. Subsequently the excitations were also shown to have
fractional statistics\cite{statistics}.

A unified understanding of all fractions $\nu=p/(2sp+1)$ was achieved
by the Composite Fermion picture of Jain\cite{jain-cf}. In this
picture, the true quasiparticles are electrons dressed by $2s$
vortices, which are called Composite Fermions (CFs). At a mean field
level, the CFs then see a reduced field $B^*=B/(2sp+1)$, in which they
fill $p$ CF-Landau levels (CF-LLs), and exhibit the integer QHE. This
picture has been very successful in obtaining excellent wave
functions\cite{jain-cf-review}. In the past few years, the
experimental reality of CFs has also been firmly
established\cite{stormer-cf}.

Great progress has also been made in arriving at a functional
integral description in which some of the nontrivial properties found in the
wave-function approach arise at the mean field level. The functional
treatment is based on the Chern-Simons(CS) transformation\cite{cs-trans}, which
performs flux attachment via the CS gauge field to obtain either
bosons\cite{gcs,gmcs,zhk,read1} or fermions\cite{lopez}. These
theories have provided us with a link between the microscopic
formulation of the problem and experiment, both for incompressible and
compressible states\cite{kalmeyer,hlr}.

Recently R. Shankar and the present author developed a hamiltonian CS
theory for the FQH states
\cite{us1,us2}. Inspired by  the work
of Bohm and Pines\cite{bohm-pines} on the 3D electron gas, we enlarged
the Hilbert space to introduce $n$ high-energy magnetoplasmons degrees
of freedom, ($n$ also being the number of electrons) at the same time
imposing an equal number of constraints on physical states.  Upon
ignoring the coupling between the oscillators and the fermions we
obtained some well known wavefunctions
\cite{laugh,jain-cf}. However the fermions still had the bare mass,
and the frequency of the magnetoplasmons was incorrect. Hence a final
canonical transformation was employed to decouple the fermions from
the oscillators {\em in the infrared limit}.

	We choose to call the final fermions {\em the} composite
fermions for the following reasons.  Firstly, the final fermions have
no dispersion in the absence of interactions and acquire an effective
mass dependent on interactions alone.  Next, the final canonical
transformation assigns to each fermion the magnetic moment $e/2m$ as
mandated by the arguments of Refs.
\cite{ssh,simon-f1}. The central result of our formalism is the formula for the
electronic charge density, which takes the following form, separable
into high- and low-energy pieces\cite{us2}, at small $q$:
\eqa
\rho_e(q)={q \over \sqrt{8\pi}} \sqrt{{2p\over2p+1}} (A(q)+
A^{\dagger}
(-q))&\nonumber\\
+{\sum_j e^{-iqx_j} \over 2p+1} -{il_{}^{2} }  (\sum_j (q \times
\Pi_j)e^{-iqx_j}&
)\label{rhobar}
\eea
where $A,A^{\dagger}$ refer to the annihilation and creation operators
of the magnetoplasmon oscillators, $l =1/\sqrt{eB}$ is the magnetic
length, and ${\vec\Pi}_j={\vec P}_j+e{\vec A}^*(r_j)$ is the velocity
operator of the CFs.  The oscillator piece saturates Kohn's
theorem\cite{kohn}. The rest, to be called $\rhob$, is obtained by
adding to the canonically transformed electronic charge density a
particular multiple of the constraint\cite{us2} (in the physical
subspace, one can add any multiple of the constraint without physical
consequences, but we wish to work in the full space). It has some very
useful properties in the full space:

\begin{itemize}
\item{} $\rhob$ satisfies the
magnetic translation algebra (MTA)\cite{GMP} to lowest leading
order. Since this is the algebra of the electron density in the lowest
Landau level (LLL), we have performed the LLL projection correctly in
the infrared.
\item{} Note that $\rhob$ is a sum of a monopole with charge
$e^*=e/(2p+1)$, which is the charge associated with the CF, and a
dipole piece which alone survives at $\nu=1/2$ and has the value
proposed by Read\cite{read2}.  (A number of recent constructions have
emphasized this dipolar aspect\cite{dh,pasquier,read3}). 
\item{} We also find
that that as $\vq\to 0$ all transition matrix elements of $\rhob$ from
the HF ground state vanish at least as $q^2$.
\end{itemize}

The final property is an essential property of {\it physical } charge
density matrix elements from incompressible liquid ground states in the
LLL\cite{GMP}.  We will present arguments in the next section to show that if
one intends to use the Hartree-Fock approximation ignoring
constraints, these properties of $\rhob$ are essential. They make it
plausible that $\rhob$ does not suffer vertex corrections.

 The Hamiltonian of the low-energy sector  (dropping the magnetic
moment term)
is
\eq
H=\half \int {d^2 q\over(2\pi)^2} v(q) \rhob(-q) \rhob(q)\label{ham}.
\ee
where $v(q)$ is the electron-electron interaction.  We circumvent the
  fact that $\bar{\rho}$ is to be trusted only for small $q$ as
  follows.  Consider real samples which have a finite thickness
  $\Lambda$ of the same order as $l $, so that the Coulomb interaction
  is cutoff at large wavevectors\cite{thick1}. It was realized by
  Haldane and Rezayi\cite{hal-rez} that this has a large effect on the
  gap, while leaving the wavefunctions essentially unchanged. We will
  focus on such interactions, parametrized by $\lam =\Lambda /l $ for
  which numerical results for the transport gaps are
  available\cite{thick2,bonesteel,jain-th}. The advantage is that as
  $\lam$ becomes large only small-$q$ matrix elements of the density
  are invoked in computing gaps, and we expect our theory to become
  more accurate. It is possible that beyond some large $\lam$ the
  incompressible liquid might cease to be the ground
  state\cite{hal-rez}. Our theory, which is based on a liquid state
  with uniform density, can be expected to work up to this $\lam$. In
  fact, the magentoexciton dispersions can be used to infer these
  instabilities, as will be shown later.

This hamiltonian is to be supplemented by $n$ constraints which
identify the physical subspace. We will expand on this issue in the
next section.  In two earlier papers we presented calculations of gaps
for a few fractions\cite{single-part} in the Hartree-Fock
approximation, and tested certain scaling relations that arise
naturally in the our theory against numerical results obtained using
CF wave functions\cite{scaling}. Emboldened by the good agreement we
found between the predictions of our theory and numerical results, in
this paper I present results for magnetoexciton (ME) dispersions for
the polarized FQHE states at $\nu=\third$, $\twof$, and $\threes$
based on our formalism. I will also present results for the spin-flip
MEs, one branch (not always the lowest) of which corresponds to the
spin waves in the ferromagnetic polarized states.

The computation of ME dispersions has a long history\cite{me-history}
in the IQHE. Many different approximations\cite{me-history,kallin} have been
employed to obtain the ME dispersions in the IQHE, most of which can
be subsumed into the powerful time-dependent HF (TDHF) treatment of
MacDonald\cite{macd1,macd2}. In this approach one allows the
hamiltonian to scatter particle-hole excitations with different
Landau-level indices into each other, and diagonalizes the resulting
matrix.

A serious problem arises when considering ME dispersions in the
FQHE. Since the states are strongly correlated in terms of electron
variables, it is difficult to find good approximation
techniques. Consequently, the most trusted results are those of exact
diagonalizations on finite systems\cite{hal-rez,exact-ex}. There are also
results from other methods, such as the single-mode approximation
(SMA)\cite{GMP}, which works well near the roton-like minimum for
$\nu=\third$, and the use of CF wavefunctions for excited
states\cite{jain-ex}. Field-theoretic approaches include the seminal
work of Lopez and Fradkin\cite{lopez} for the collective modes of
gapped fractions based on a fermionic CS approach, and an RPA
treatment\cite{rpa-ex} based on the formalism of Halperin, Lee, and
Read (HLR)\cite{hlr}. These field-theoretic approaches suffer from the
problem that the bare mass (actually the band mass $m_b$)\cite{lopez}
or a phemenological mass $m^*$ enters the dispersions, whereas in the
LLL the scale should be set solely by
$e^2/\varepsilon l$.

Our approach essentially solves the problem of strong correlation by
rewriting the theory in terms of CFs, which are the quasiparticles of
the theory in the same sense as those in Landau's Fermi Liquid theory:
The quasiparticles can interact strongly with a Fermi-Liquid-like
interaction, but matrix elements which scatter them out of their
states are small at low energies, allowing them to be
long-lived. Furthermore, the only scale in our low-energy hamiltonian
is indeed $e^2/\varepsilon l$, as desired. For the reasons to be explained below,
we {\it can} use a formalism such as TDHF, which ignores vertex
corrections, with our formula for $\rhob$. 

It must be emphasized that the TDHF approach of
MacDonald\cite{macd1,macd2} which includes many Landau levels (LLs) is
absolutely essential for treating the FQHE case. In the integer QHE
one has two distinct energy scales, $\hbar\o_c$ and $e^2/\varepsilon
l$, and one may go to the strong-field limit\cite{kallin}
$\hbar\o_c>>e^2/\varepsilon l$ to make the problem theoretically
convenient without losing any essential physics. In this limit only
the lowest unoccupied and highest occupied LL need be considered, and
LL-mixing of the MEs can be ignored. However, in our LLL analysis of
the FQHE, there is only one energy scale, and the variation of energy
within an ME mode is comparable to the splitting between different
modes. We will see that the correct magnetoroton minimum cannot be
obtained without considering this ``CF-LL mixing''.

The spin quantum number produces additional nontrivial features. It
was realized by Sondhi {\it et al}\cite{shivaji-skyrmion} that even in
the absence of Zeeman interactions, the $\nu=1$ state, and
correspondingly the Laughlin fractions $1/2p+1$, were ferromagnetic
due to exchange interactions. The symmetry-breaking leads to gapless
spin waves with a quadratic dispersion. If the Zeeman interaction is
now turned on the spin waves acquire a gap. Due to the fact that all
other excitations of the system are much higher in energy than the
spin waves, the $\nu=1$ and the Laughlin fractions are dynamically
identical to a simple quantum ferromagnet at low
temperatures\cite{q-fm-th}. The detailed predictions have been
dramatically confirmed by experimental results\cite{q-fm-ex}. There
have been few attempts to compute the low-energy spin-flip modes for
the polarized FQH states. One, due to Nakajima and
Aoki\cite{nakajima}, relates the FQH spin-wave mode to the IQH
spin-wave mode, performing the flux-attachment by shifting the
pseudopotentials. This, when tested against exact
diagonalizations\cite{nakajima}, produces good results for the
$\nu=\third$ and ${1\over5}$ spin-wave modes. Another approach, by
Mandal\cite{mandal}, uses fermionic CS theory to compute the response
functions, and from them the spin-flip dispersions. In the spirit of
the earlier CS approaches\cite{lopez,rpa-ex} a phenomenological
effective mass is used to approximate the energies of the CF-LLs, and
the strong-field approximation is used. Where appropriate, we will
comment on the relationship between these and our results.

The plan of the paper is as follows.  In section II we will set up the
conventions and define various terms. We will also show that the
states with $p$ filled CF-LLs are indeed HF ground states for
$\nu=p/2p+1$, and we will make some important remarks about the
applicability of TDHF in our approximation scheme for the gapped
fractions. In section III we describe the method of computing the ME
dispersions, which is essentially MacDonald's approach\cite{macd1}
expressed in an operator form. In Section IV we present the results
for the spin-polarized and the spin-flip ME dispersions and compare
them to previous work. As mentioned earlier, beyond some $\Lambda$ the
incompressible FQHE state becomes unstable\cite{hal-rez}. Other
instabilities include the transition to a Wigner crystal at small
filling\cite{wc-instability}. The latter instability has been studied
by looking at the ME dispersion in the $1/7$\cite{GMP,jain-wc} and
$1/9$ systems\cite{jain-wc}. It is found that the ME dispersion
becomes negative at finite wavevector\cite{jain-wc} for the $1/9$ FQHE
state, indicating that the ground state is unstable to the creation of
large numbers of MEs. In analogy with this work, we will go back to
the $\third$, $\twof$, and $\threes$ FQHE states at large $\Lambda$
and find instabilities in the ME dispersions for $\nu=\twof$ and
$\threes$.  We end with conclusions and open questions in section V.

\section{Setting up the Formalism}
\label{sec2}

Let us return to the hamiltonian of equation (2), focussing for
simplicity on spin-polarized states.  Since we have a many-body
Hamiltonian, we should choose a many-body ground state (GS). The
formula for $\rhob$ contains the velocity operator for the CFs in the
effective field $B^*=B/2p+1$, suggesting that the CFs fill CF-LLs.

We now have a problem where the states are the same as those for
(spinless) fermions in the IQHE.  We work in the symmetric gauge, in
which the single-particle wavefunctions are characterized by the LL
index $n$ and the angular momentum index $k$.  The formalism in this
gauge has been extensively developed\cite{laugh,fetter} in the context
of anyon superconductivity. To conform to the notations of
\cite{fetter} we will define the magnetic field to be along the {\it
positive} $\hat{z}$ direction, which produces the wave functions

\eq
\phi_{n,k}(\vrr) = {1\over\sqrt{2\pi 2^{n+k}n!k!}}({z\over2}-2\partial^*)^n 
({z^*\over2}-2\partial)^k e^{-{|z|^2/4}}
\ee
Here $z=x+iy$, and all distances are measured in units of the
effective cyclotron length $l^*=l\sqrt{2p+1}$. Note that the LLL wave
function is {\it antianalytic} with this choice of field direction.

The density is a one-body operator, and can create particle-hole pairs
above the GS. It is well-known\cite{lerner} that these pairs come with
a conserved momentum $\vq$, which also characterizes the separation
between the particle and the hole $\vrr_{ph}={\hat z}\times\vq
l^{*2}$.  The magnetoexciton wave functions\cite{lerner}
$\psi_{nn'}(q;\vrr,\vrr')$ describe a particle in the $n^{th}$ LL
and a hole in the $n'^{th}$ LL. In the notation
of ref.\cite{fetter}, it is expressed as

\eqa
\psi_{nn'}(\vq;\vrr,\vrr')=&{(-1)^n\over L}{1\over\sqrt{2\pi 2^{n+n'}n!n'!}}
(2\partial_1^*-{z_1\over2})^n (2\partial_2-{z_2^*\over2})^{n'}\nonumber\\
&e^{-{1\over4}(r_1^2+r_2^2+Q^2)} e^{\half(z_1^*z_2+z_1^*z_q-z_q^*z_2)}
\eea
Here $L$ is the linear size of the system in units of $l^*$ (the
dimensionless area is $L^2$), and $z_q=iQ_+=il^*(q_x+iq_y)$ is the
dimensionless wavevector.

We can now define the $x$-coefficients, which denote the overlap
between particular single-particle + single-hole states and the ME
wavefunction as (for $k\ge k'$)

\eqa
x_{nn'}^{kk'}(q)&=\int d^2\vrr_1 d^2\vrr_2 \phi_{nk}^*(\vrr_1) \phi_{n'k'}(\vrr_2)
\psi_{nn'}(\vq:\vrr_1,\vrr_2)\nonumber\\
&=
(-1)^{n'}{\sqrt{2\pi}\over L}\sqrt{{k'!\over
k!}}\bigl({iQ_+\over \sqrt{2}}\bigr)^{k-k'} L_{k'}^{k-k'}(y)e^{-y/2}
\label{xco}\eea
Here $L_n^m$ is a Laguerre polynomial, and $y=Q^2/2$. The
$x$-coefficients are also the matrix elements of the unitary
transformation that relates the product particle-hole basis with the
ME basis. In order to obtain the $x$-coefficients for $k<k'$ we use 
\eq
x_{n'n}^{k'k}(q)=(-1)^{n+n'}(x_{nn'}^{kk'}(-q))^*
\ee

 Now we are
ready to express $\rhob$ in this basis set. The form of $\rhob$
presented in Eq.(\ref{rhobar}) is correct for the field direction
being $-\hat{z}$. In order to conform to the notation we are
using\cite{fetter} we have to use the following form
\eq
\rhob(q)={\sum_j e^{-iqx_j} \over 2p+1} +{il_{}^{2} }  (\sum_j (q \times
\Pi_j)e^{-iqx_j}
)\label{rhobardai}
\ee
 where $\Pi_j$ is now the velocity operator in the effective field
pointing in the $+\hat{z}$ direction. Now we invert
Eq. (\ref{xco}) to get

\eq
\phi_{\nu_1}^*(\vrr_1)\phi_{\nu_2}(\vrr_2)=\intq (x_{\nu_2\nu_1}(q))^* \psi_{n_2n_1}(\vq;\vrr_2,\vrr_1)
\ee

where we have used the compact notation $\nu_i=(n_i,k_i)$. To express the density in terms of second quantized operators we note that 
\eqa
\hat{\rhob}(q)=&\int d^2r \bigg({e^{-iqr}\over2p+1}+il^2 q\times\Pi e^{-iqr}\bigg) d^\dagger(\vrr) d(\vrr)\nonumber\\
=&\sum\limits_{\nu_1\nu_2}^{} \rho_{n_2n_1}(\vq)(x_{\nu_2\nu_1}(\vq))^* d^{\dagger}_{\nu_1}d_{\nu_2}
\eea
where we define 
\eqa
\rho_{n_2n_1}(\vq)=&\int d^2r \psi_{n_2n_1}(q;\vrr,\vrr)\bigg({e^{-iqr}\over2p+1}+il^2 q\times\Pi e^{-iqr}\bigg)\nonumber\\
=&{(-1)^{n_1}\over
2p+1}{L\over\sqrt{2\pi}}\sqrt{n_1!\over
n_2!}\bigl({-iQ_+\over\sqrt{2}}\bigr)^{n_2-n_1}\nonumber \\
\times
&e^{-y/2}(n_2L_{n_1-1}^{n_2-n_1}+2L_{n_1}^{n_2-n_1}-(n_1+1)L_{n_1+1}^{n_2-n_1})
\label{rhomat1}\eea
 This last expression is valid for $n_2\ge n_1$. For $n_2<n_1$ one uses the relation 
\eq 
\rho_{n_2n_1}(\vq)=(-1)^{n_1+n_2}(\rho_{n_1n_2}(-\vq))^*
\label{rhomat2}\ee 
In what follows, it will be useful to define the magnetoexciton operators
\eq
\hX_{n_1n_2}(\vq)=\sum\limits_{k_1k_2} (x_{\nu_2\nu_1}(\vq))^* d^{\dagger}_{\nu_1}d_{\nu_2}
\label{meops}\ee
where the sums are over the angular momentum indices. Finally, one can compactly express the electron density operator as 
\eq
{\hat\rhob}(\vq)=\sum\limits_{n_1n_2} \rho_{n_2n_1}(\vq)\hX_{n_1n_2}(\vq)\label{rhob}
\ee

Apart from the trivial dependence of $x$ on $n'$, we see the
separation between the angular $k$ labels and the ``radial'' $n$
labels. 

We will need the following identities, which can be easily established
by using the defining relation for $x$, Eq. (\ref{xco});
\eqa
\sum\limits_{k}^{}x_{nn}^{kk}(\vq) =(-1)^n
&{L\over\sqrt{2\pi}}&\delta_{{\vec Q},0}\\
\sum\limits_{k}^{}
x_{n_1n}^{k_1k}(\vq_1)x_{nn_2}^{kk_2}(\vq_2)&=(-1)^n&{\sqrt{2\pi}\over 
 L}e^{{-i\over2}\vQ_1\times{\vQ}_2}\times\nonumber\\
&&x_{n_1n_2}^{k_1k_2}(\vq_1+\vq_2)\label{identities}
\eea

\subsection{Calculation of Gaps}

As mentioned before, we have to assume a many-body ground state. Since
the low-energy hamiltonian contains the velocity operator in the
effective field $B^*=B/2p+1$, it is natural to assume a ground state
of $p$ filled CF-LLs for $\nu=p/2p+1$. This is not an eigenstate of
$H$, which can create particle-hole pairs above it. However, it is the
HF ground state. We now establish that, for rotationally invariant
interactions, the Hamiltonian does not mix a single-particle (or
single-hole) state with any other single-particle (or single-hole)
state. This is the signature of a HF state.

We begin by writing down the hamiltonian in terms of the CF creation and annihilation operators
\eqa
H=&\intq{v(q)\over2} \sum\limits_{\nu_i} x_{\nu_4\nu_1}^*(-\vq)x_{\nu_3\nu_2}^*(\vq) \nonumber\\
\times&\rho_{n_4n_1}(-\vq)\rho_{n_3n_2}(\vq)\dd_{\nu_1}d_{\nu_4}\dd_{\nu_2}d_{\nu_3}
\eea

Consider the transition matrix element of $H$ between two {\it
different} single-particle states $\mu=(m,k)\ne (m',k')=\mu'$ on top
of the assumed GS of $p$ filled CF-LLs;

\eqa
<GS|d_{\mu}H\dd_{\mu'}|GS>=\half \intq
v(q)\sum\limits_{\nu_i}& \nonumber\\
\rho_{n_4n_1}(-\vq)\rho_{n_3n_2}(\vq)
x_{\nu_4\nu_1}^{*}(-\vq)x_{\nu_3\nu_2}^{*}(\vq)&\nonumber\\
<GS|d_{\mu}\dd_{\nu_1}d_{\nu_4}\dd_{\nu_2}d_{\nu_3}\dd_{\mu'}|GS>&
\eea

Now we perform the usual Wick contractions (at equal time) with 
\eqa
&<GS|\dd_{\nu} d_{\nu'}|GS>=\delta_{\nu\nu'}N_F(\nu)\\
&<GS|d_{\nu} \dd_{\nu'}|GS>=\delta_{\nu\nu'} (1-N_F(\nu))
\eea

where $N_F(\nu)$ denotes the occupation of the single-particle state
$\nu$. In performing the Wick contractions one should take care not to
contract the two operators belonging to the same density (say
$\dd_{\nu_1}$ with $d_{\nu_4}$), since this leads to the $\vq=0$
density, which is cancelled by the background positive charge. Now we
can write the transition matrix element as

\eqa
&(1-N_F(\mu))(1-N_F(\mu'))\intq{v(q)\over2} \sum\limits_{\nu}\bigg(\nonumber\\
&(1-N_F(\nu))x_{\nu\mu}^*(-\vq)x_{\mu'\nu}^*(\vq)\rho_{nm}(-\vq)\rho_{m'n}(\vq)-\nonumber\\
&N_F(\nu) x_{\mu'\nu}^*(-\vq)x_{\nu\mu}^*(\vq)\rho_{m'n}(-\vq)\rho_{nm}(\vq)\bigg)  
\eea

We now use the $x$-identities to perform the sum over the angular momentum index of $\nu$  in the two
terms to get a factor of $x_{\mu\mu'}^*(\vq=0)$ in both terms. The
explicit form of $x$ tells us that setting $\vq=0$ forces $k=k'$. This
leaves the integral over the reduced density matrix elements. Consider
the first term  for specificity
\eq
H_1={\pi\delta_{kk'}\over L^2} \int {d^2 q\over(2\pi)^2}v(q) \sum\limits_{n\ge p}
(-1)^{m+n}\rho_{mn}(-q)\rho_{nm'}(q) 
\ee

By using the explicit functional form of the density matrix elements,
Eqs. (\ref{rhomat1}) and (\ref{rhomat2}), we can reduce this to the form
\eq
H_1={\pi\delta_{kk'}\over L^2}\int {d^2 q\over(2\pi)^2}v(q) \sum\limits_{n\ge p} f(m,m',n,q^2) e^{i\theta(m-m')}
\ee
where $\theta$ is the angle of $\vq$ with respect to the $q_x$ axis. 
It is then clear that for $m\ne m'$, the rotational invariance of
$v(q)$ will force this term  to be zero. The same
result can be verified for the second term, and also for the case of two
different single-hole states. 

This establishes the HF nature of the GS, and supports our view that
we are working with the right variables. We can now verify from
Eq. (\ref{rhomat1}), as claimed in the introduction, that as $\vq\to 0$
all transition matrix elements from the HF ground state vanish at
least as $q^2$, an essential property of charge density matrix
elements in the LLL\cite{GMP}.

Let us now  find the energy of a single-particle state
above the filled sea. This is just the diagonal matrix element of $H$,
and we have already done most of the work. Setting $m=m'$ in the
above, we get

\eq
\epsilon(m)={\pi\over L^2}\intq v(q)\sum\limits_{n}(1-2N_F(n))|\rho_{nm}(\vq)|^2
\ee

The first term is the Hartree energy while the second represents the exchange, or Fock part. The hamiltonian can be normal ordered to make this obvious in operator form
\eqa
H=&H_0+V\nonumber\\
H_0=&\sum\limits_{\nu}\e_0(n)\dd_{\nu}d_{\nu} \label{h0}\\
V=&\intq{v(q)\over2} \sum\limits_{\nu_i} x_{\nu_4\nu_1}^*(-\vq)x_{\nu_3\nu_2}^*(\vq) \nonumber\\
\times&\rho_{n_4n_1}(-\vq)\rho_{n_3n_2}(\vq)\dd_{\nu_1}\dd_{\nu_2}d_{\nu_3}d_{\nu_4}\label{hhf}\\
\e_0(n)=&{\pi\over L^2}\intq v(q)\sum\limits_{n}|\rho_{nm}(\vq)|^2\label{eh}
\eea

Similarly, one can easily derive the hole energy, which is identical
to the above, except for a minus sign. Notice that the energies depend
only on the CF-LL index, and not on the angular momentum, and that
there is a sum over an infinite tower of CF-LLs in the energy.

So far our discussion is applicable to arbitrary $v(q)$. Below we will
find it convenient to specialize to a model potential that
approximates the effects of sample thickness
$v(r)=e^2/\varepsilon\sqrt{r^2+\Lambda^2}$, whose Fourier transform is
$v(q)=2\pi e^2\exp{(-q \Lambda)}/\varepsilon q$. The parameter
$\Lambda$ is related to the sample thickness.

At this point, two important questions arise: We have treated the
hamiltonian in the HF approximation, and we have ignored constraints
altogether. It is important to understand whether these
approximations are reasonable, and how they might (at least in principle) be
improved. The next four  subsections will address these issues in turn. 

\subsection{Why is Hartree-Fock reasonable?}

Hartree-Fock is an approach that works best when correlations are
weak. The ground state is sought to be expressed in terms of a single
determinant of single-particle wave functions. Clearly, carrying out a
HF calculation in terms of {\it electron} coordinates would be
nonsensical in the FQHE problem: This is the message that Laughlin's
work\cite{laugh} has driven home. However, what we are doing here is
radically different. We are working in terms of CF coordinates. The
correlated ground state of electrons can be viewed as the ground state
of independent CFs, which is the message of Jain's
work\cite{jain-cf-review}. In other words, the system of electrons
reorganizes itself into a state where the CFs are the
Fermi-liquid-like quasiparticles. Thus one can reasonably expect to
get good results from a simple approximation such as HF when working
in CF coordinates.

However, one other condition has to be satisfied if the results are to
be numerically accurate. We have to make sure that vertex corrections
are small. Large vertex corrections imply that the particle we started
out with is not the physically observed quasiparticle, {\it i.e.},
that the observed charge, dipole moment, etc are different from those
entering the microscopic hamiltonian. In this case, the HF excited
states will have small overlap with the true excited states, and the
HF energies will be correspondingly poor. Our formula for $\rhob$ does
have the correct charge and dipole moments of the final physical
particles. In other words, they are already fully dressed, and
therfore vertex corrections are expected to be small.

There is another way to conclude that vertex corrections are small for
$\rhob$: Recall\cite{us2} that in order to get $\rhob$ we took a
particular multiple of the constraint and added it to the ``naive''
transformed electron density. In principle, one can add any multiple
of the constraint to the density, and the physical matrix elements
should be unaffected. However, this particular combination has a very
nice property which we have mentioned before: All its transition
matrix elements from the HF ground state are order $q^2$ or higher.
This is a necessary property of all {\it physical} density matrix
elements in the incompressible fractions\cite{GMP}. This property does
not hold for the naive transformed density, or for any other linear
combination of the density and the constraint; they all have
transition matrix elements of order $q$ from the assumed HF ground
state. Of course, if the constraint were exactly implemented, it would
force all physical matrix elements to behave correctly by producing
vertex corrections.  Thus, every combination except our $\rhob$ is
{\it guaranteed} to suffer strong vertex corrections. Therefore we
come to an important conclusion: If we intend to use HF without
implementing the constraint exactly, we must use $\rhob$, and only
$\rhob$, for the electron density operator.

\subsection{Constraints}

Since it is very hard to implement these constraints exactly, we will
use two different approaches. The first approach is to ignore the
constraints altogether. The rationale for doing this is the following:
Suppose we had the exact expression for the electron density operator
in final coordinates. Since this operator is gauge invariant ({\it
i.e.} commutes with the constraints), it will have no matrix elements
between the true physical ground state and unphysical states. Thus the
sum over physical states can be extended to a sum over all states with
no problem. However, this requires that we also have the correct
physical ground state. In practice we have neither an exact expression
for $\rho_e$ nor the true ground state. However, our
$\rhob$ is close to an operator that is  gauge invariant\cite{aftermath} (we will elucidate this in subsection E), and we can hope that
the ground state we choose is close to the true one, so that all the
above statements are approximately correct.

While this approach indeed produces reasonable answers for the
thickness parameter $\lam \ge 2.0$ or so, it produces an unphysical
divergence of the transport gap for $\lam=0$\cite{single-part}. The
reason is not far to seek. For small $\lam$ the potential has
significant weight at large $q$, where our expression for $\rhob$ is
not to be trusted. Thus in this region, our $\rhob$ couples to
unphysical states, of which there are an infinite number, which
produces the divergence. Thus, in order to get a physically sensible
answer for small $\lam$ one is forced to account for the constraint in
some way. We choose the simplest possible way, by cutting off the high
energy CF-LL states. There are many ways of seeing that such a cutoff
makes physical sense. Firstly, for a finite number of particles, there
are a finite number of states in the many-body Hilbert space (after
projecting to the LLL, which we do by freezing oscillators),
equivalent in dimensionality to the Hilbert space generated by $2p+1$
CF-LLs. Secondly, density matrix elements in the LLL are suppressed at
high momenta by gaussian factors. Matrix-elements of our $\rhob$
between higher CF-LL states are peaked at higher momenta, and thus it
is plausible that the higher CF-LLs are less physical. Finally, even
in the LLL, we expect CFs to describe only the low-energy dynamics. At
energies of the order of the first Haldane
pseudopotential\cite{haldane-book} $V_1$, we expect the CF to break up
into electrons and vortices. Thus, CF-LLs cannot be the correct
description at high energies. On the other hand, there is ample
numerical evidence that the low-lying CF-LLs are very close to true
physical states\cite{dev}.

Thus we have the following picture: The higher CF-LLs probably have
very little overlap with physical states, while the lower ones have
large overlap. This argument does not tell us precisely which states
we must keep. In the rest of this paper I will choose a sharp cutoff
in the CF-LLs, wherein some number (around $2p+1$) of the lowest ones
are kept, and the rest discarded. We will also investigate how
sensitive our results are to a change in this cutoff, and we will find
that for $\lam\ge2.0$ or so, the results become insensitive, and that
one can keep all the CF-LLs without much altering the results. The
main operational aim of this method of implementing the constraints is
to obtain qualitatively reasonable (though perhaps numerically
untrustworthy) results for small $\lam$.

Thus far we have given arguments mostly from within our theory. Of
course, the correct way to settle matters of principle is to implement
the constraints from the start on an equal footing with the
hamiltonian. Such a calculation has not been carried out for the
gapped fractions, but Read\cite{read3} has recently studied the case
of $\nu=1$ bosons in detail (this belongs to class of compressible
systems, among which is included the half-filled LLL for
electrons\cite{kalmeyer,hlr}), where constraints are crucial in
restoring compressibility to the
system\cite{hlr,dh,read3,comment,stern}. We devote the next subsection
to the lessons we can draw from Read's work\cite{read3}.

\subsection{Read's Conserving Approximation for $\nu=1$ Bosons}

Read\cite{read3} starts from the LLL formulation of Pasquier and
Haldane\cite{pasquier} for $\nu=1$ bosons, and implements the
constraints (which, unlike the constraints in our
formulation\cite{us1,us2}, do not commute among themselves, but form
an algebra) in a path-integral conserving
approximation\cite{conserving}. Our greatest interest is in the gaps
and ME dispersions, which can be read off as the poles of the
density-density correlator. Let us therefore focus on his results for
the irreducible density correlator $\chi_{irr}$, for which he
finds\cite{read3} that:

\begin{itemize}
\item{} $\chi_{irr}$ breaks up naturally into two additive 
contributions. The first is the density correlator of dipoles, while
the second represents the dipoles interacting with a transverse gauge
field. It should be emphasized that the fermions that enter the 
lagrangian have charge $e$, and that a nontrivial renormalization has
happened due to vertex corrections in order to make them dipolar. 

\item{} The contribution due to the transverse gauge field 
dominates at very low frequency, restoring compressibility to the
system. This contribution controls the $\omega^0$ and the
$\omega^{-1}$ moments of the density spectral function.

\item{} At higher frequencies, the purely dipolar part dominates. 
For example, the $\omega$ moment of the density spectral function is
identical to the result for purely dipolar fermions.
\end{itemize} 
Let us view our method in the light of these results. Our fermionic
density, written in terms of $\rhob$, is dipolar (for $\nu=\half$)
already at tree level, and no further vertex corrections are
necessary. What is missing in our approach is the coupling to the
low-energy transverse gauge field. However, this contribution seems to
be relatively unimportant at high frequencies.

While we are not certain how all this might extend to the gapped
fractions, we can make the following plausible guesses: The vertex
corrections will renormalize the charge of the fermion from $e$ to
$e^*$, and produce a dipole moment (and perhaps higher multipole
moments). Once again this renormalization (upto dipole terms) is
present in our $\rhob$ already at tree level. Secondly, we can expect
a gauge field to enter the picture. However, we once again expect it
to be important only at low frequencies.  Since all the answers we
seek in the gapped fractions are at finite frequencies of the order of
the gaps, we do not expect the gauge field to make a large
contribution to our results.

To summarize, the effects of the constraint in Read's calculation are
twofold: They renormalize the charge and dipole moment, and they
create a transverse gauge field to which the dipolar fermions are
coupled, and which is important at low frequencies. The first effect
is present in full in our approach, while we argue that for the
physical quantities of interest to us, the contribution of the second
can be safely ignored in the gapped fractions. 

We now turn to a very recent new formulation for the gapped fractions.

\subsection{Shankar's New Formulation for Principal Fractions}

Very recently, Shankar\cite{aftermath} has proposed formulas for the
charge density and constraint for all principal fractions. He writes
both the electronic charge density (call it $\rhob_S$) and the
constraint (call it $\chi_S$) as a sum over exponentials of single-CF
coordinates. The first two terms of each exponential are exactly the
two terms derived in our earlier work\cite{us2}. $\rhob_S$ and
$\chi_S$ have some very nice properties:

\begin{itemize} 
\item{} $\rhob_S$ satisfies the MTA {\it to all orders} in $q$. 
This formulation is not restricted to the infrared.
\item{} $\chi_S$ also satisfies the MTA with a modified magnetic 
length, and is identified as the vortex charge
density\cite{aftermath}. (It should be noted that this is different
from our original formulation, since our constraints
commute\cite{us1,us2}, and should continue to do so after the
canonical transformations).
\item{} $\rhob_S$ and $\chi_S$ commute with each other, so $\rhob_S$ 
is gauge invariant.
\item{} The hamiltonian can be written in terms of $\rhob_S$, 
which makes it  gauge invariant, and amenable to a conserving
approximation.
\end{itemize}

Such a conserving approximation would be the way to settle matters of
principle, such as the quantum numbers of the excitations in the
gapped fractions, and their statistics\cite{statistics}. However,
Shankar also points out\cite{aftermath} that there is another possible
formulation of the problem that is very closely related to that
followed in this paper, and provides a justification of it. He notes
that one can form a preferred charge density
\eq
\rhob_{pref}=\rhob_S+c\chi_S
\ee
where $c$ is the charge of the vortex. This charge density has a very
appealing physical interpretation, since each composite fermion has an
electron and a vortex in close proximity, and $\rhob_{pref}$ is just
the total charge of such a composite. One can now express the
hamiltonian in terms of $\rhob_{pref}$, and it commutes weakly with
the constraints ($[H,\chi]\simeq \chi$). In the physical subspace this
is as good a formulation as the one in the previous paragraph.

Most important for our purposes, the first two terms of this preferred
charge density are exactly the two terms in our $\rhob$,
Eq. (\ref{rhobar}). Therefore $\rhob_{pref}$ enjoys all the nice
properties of $\rhob$, such as explicitly displaying the right charge
and dipole moment, satisfying the MTA at lowest leading order, and
having $q^2$ matrix elements out of the HF ground state. Shankar
points out\cite{aftermath} that if one is interested in computing
numbers in the gapped fractions, this second formulation, with many
nontrivial renormalizations built in at tree level, may produce better
numbers than the conserving approach.

This second viewpoint meshes perfectly with our philosophy. It also
gives us an algebraically consistent way to continue the expression
for $\rhob$ to all orders in $q$. When one looks at the corrections to
our $\rhob$ by comparing it to Shankar's $\rhob_{pref}$, one finds
that the leading order correction (order $q^2$) vanishes, and the next
correction (order $q^3$) is down by a large factor
($\approx500$)\cite{aftermath}. This suggests that while our $\rhob$
was derived at small $q$, it regime of validity may be larger than we
have any right to expect.

To summarize the past four subsections, we have presented several
lines of reasoning that suggest that by using our $\rhob$ for the
electronic charge density, we are working in a preferred
representation where the effect of constraints is minimal, and
approximations such as HF and TDHF can be expected to give good
answers. We now proceed to the calculations.

\section{Method of Calculation}
\label{sec3}

Conventionally, in using the TDHF approximation\cite{macd1}, one first
computes the density matrix under a time-dependent perturbation, and
thence the behavior of the density with time to linear order in the
external potential. The susceptibility is directly connected to the
density-density correlator, and its poles give the ME dispersions.

We will be following an operator variant of this method, where we will
directly determine the time dependence of the ME operators (defined in
Eq.  (\ref{meops})) by commuting them with the hamiltonian. This
method also has an ancient history\cite{operator-rpa}. The expression
on the right hand side will, in the spirit of HF, be simplified by
reducing four-fermi operators to two-fermi ones by taking averages.

We will initially consider the simpler spin-polarized case, and later
show how to extend the formalism to include spin. 
\medskip
\subsection{Spin-Polarized Magnetoexcitons}
\medskip

For a Heisenberg
operator we have the relation
\eq
{\p O\over \p t}=i[H,O]
\ee
Let us begin by considering the time dependence of the ME operator
$\hX_{m_1m_2}(q)$ by computing $[\hH,\hX_{m_1m_2}(q)]$. We will find
it convenient to use the form of $\hH$ split into Hartree and
interaction terms as given in Eqs. (\ref{h0},\ref{hhf}). It is then
easy to verify that
\eq
[\hH_0,\hX_{m_1m_2}(\vq)]=(\e_0(m_1)-\e_0(m_2))\hX_{m_1m_2}(\vq)
\ee
where $\e_0$ is just the Hartree energy of Eq. (\ref{eh}). To compute
the next term we will need
\eqa
&[\dd_{\nu_1}\dd_{\nu_2}d_{\nu_3}d_{\nu_4},\dd_{\mu_1}d_{\mu_2}]=\nonumber \\
&\d_{\mu_1\nu_4} \dd_{\nu_1}\dd_{\nu_2}d_{\nu_3}d_{\mu_2}-\d_{\mu_2\nu_1}\dd_{\mu_1}\dd_{\nu_2}d_{\nu_3}d_{\nu_4}\label{first2}\\
&-\d_{\mu_1\nu_3} \dd_{\nu_1}\dd_{\nu_2}d_{\nu_4}d_{\mu_2} + \d_{\mu_2\nu_2} \dd_{\mu_1}\dd_{\nu_1}d_{\nu_3}d_{\nu_4}
\eea

The hamiltonian has the symmetry $(\nu_1,\nu_3)\rightarrow
(\nu_2,\nu_4)$, and hence in the following we will keep only the first
two terms (Eq. (\ref{first2})), while removing the factor of $\half$ in front
of the interaction. Also, in TDHF one has to reduce all the four-point
to two-point terms. We do this by contracting one creation and one
annihilation operator in all possible ways, remembering that the
operators that belong to the same density are not allowed to contract
together (such a term would correspond to a uniform density, which is
cancelled by the positive background). These terms can be grouped as
follows:
\eqa
&N_F(m_1)\d_{\mu_2\nu_1}(\d_{\mu_1\nu_3}\dd_{\nu_2}d_{\nu_4}-\d_{\mu_1\nu_4}\dd_{\nu_2}d_{\nu_3})\label{line1}\\
&+N_F(m_2)\d_{\mu_1\nu_4}(\d_{\nu_1\mu_2}\dd_{\nu_2}d_{\nu_3}-\d_{\nu_2\mu_2}\dd_{\nu_1}d_{\nu_3}\label{line2}\\
&+N_F(n_2)\d_{\mu_2\nu_1}\d_{\nu_2\nu_4}\dd_{\mu_1}d_{\nu_3} - N_F(n_1)\d_{\mu_1\nu_4}\d_{\nu_1\nu_3}\dd_{\nu_2}d_{\mu_2}\label{line3}
\eea
Now we multiply the above by the appropriate factors and integrate
over $\vq'$, and use the $x$-identities. It is easy to show that the
last line (Eq. (\ref{line3})) produces just the Fock energies of $m_1,m_2$, which occur in the right combination to add to $[\hH_0,\hX]$ to give
\eq
(\e(m_1)-\e(m_2))\hX_{m_1m_2}(\vq)
\ee

Let us now consider the more interesting terms from Eq. (\ref{line1})
and Eq. (\ref{line2}).  There are four such terms, and they are equal
in pairs. Performing the Wick contractions and using the identities
(bearing in mind that $\delta_{\vQ,0}={(2\pi)^2\over
L^2}\delta^2(\vQ)={(2\pi)^2\over L^2(l^*)^2}\delta^2(\vq)$), we obtain
\eqa
&[\hV,X_{m_1m_2}(\vq)]=-\sul_{n_1n_2} X_{n_1n_2}(\vq)(N_F(m_1)-N_F(m_2)) \times\nonumber\\
&\bigg((-1)^{m_1+m_2}{1\over(l^*)^2 L^2} v(q) \rho_{m_1m_2}(-\vq)\rho_{n_2n_1}(\vq)\label{ed}\\
&-(-1)^{m_1+m_2}{2\pi\over L^2}\int{d^2s\over(2\pi)^2}v(s)\rho_{m_1n_1}(-\vs)\rho_{n_2m_2}(\vs)e^{i\vS\times\vQ}\bigg)\nonumber\\
&\label{ex}
\eea
where $\vQ=l^*\vq$, $\vS=l^*\vs$. For the case of the diagonal matrix
elements, {\it i.e.}, $(n_1n_2)=(m_1m_2)$, these two terms have a
simple physical meaning. The first term is the direct interaction
between the particles, which shows up as an {\it exchange} interaction
between the particle and the hole, while the second is the exchange
interaction between particles, or the {\it direct} Coulomb interaction
between the particle and hole. 
We can now write the entire commutator as 
\eqa
&[\hH,\hX_{m_1m_2}(\vq)]=(\e(m_1)-\e(m_2))\hX_{m_1m_2}(\vq)\nonumber\\
&-(N_F(m_1)-N_F(m_2))\sum\limits_{n_1n_2}(-1)^{m_1+m_2}\hX_{n_1n_2}(\vq)\times\nonumber\\
&\bigg({v(q)\over L^2l^{*2}}\rho_{m_1m_2}(-\vq)\rho_{n_2n_1}(\vq)\nonumber\\
&-\int {d^2s\over(2\pi)^2} v(s)e^{i\vS\times\vQ} {2\pi\over L^2} \rho_{m_1n_1}(-\vs)\rho_{n_2m_2}(\vs)\bigg)\label{tdhfham}
\eea
The hamiltonian can now be thought of as a matrix in the space of
``naive'' MEs, which are represented by the $\hX$ operators. These are
not eigenoperators of $\hH$, which has transition matrix elements
between different naive MEs. An important point is that both positive
and negative energy naive MEs are connected by the hamiltonian, which
is not hermitian. Nevertheless, the eigenvalues will occur in pairs
$\pm
\o_j$, with $\o_j \ge 0$ being identified as the  energy of a ME mode.
Finally, notice that both terms of an
arbitrary transition matrix element go to zero in the limit of large
$q$, the first because the matrix elements $\rho_{nn`}$ are suppressed
with gaussian factors, and the second due to the oscillatory factor
$e^{i\vQ\times\vS}$. Thus the naive MEs decouple and become the 
eigenstates of the TDHF hamiltonian in the limit of large $q$.

The density  matrix elements $\rho_{n_1n_2}$ are complex, which means that
generic Hamiltonian matrix elements are also complex. To simplify the
future discussion we define the following {\it reduced density
matrix element} $\rhot$ by the defining relation
\eq
\rho_{n_1n_2}(\vq)={L\over\sqrt{2\pi}}{1\over(2p+1)} e^{i(n_1-n_2)(\t_q-\pi/2)} \rhot_{n_1n_2}(Q)\label{rhot}
\ee
 where $\rhot$ is
real, symmetric in ${n_1n_2}$, independent of the direction of $\vq$,
and is a function of $Q=l^*q$ alone.  It is clear that the $\rhob$
matrix element can be made real by choosing $\t_q=\pi/2$, thus making
the direct matrix element of $H$, Eq. (\ref{ed}), real. The same
choice also makes the exchange matrix element, Eq. (\ref{ex}),
real.  The fact that all matrix
elements of $H$ in the ME subspace are real speeds the numerical
diagonalization.

\subsection{Scaling of ME Dispersions}
\medskip

The discussion so far is valid for an arbitrary potential. Let us now
specialize to our model potential
\eq
v(q)={2\pi e^2\over q}e^{-\Lambda q}={2\pi l^* e^2\over Q}e^{-\lam^* Q} 
\ee
where $\lam^*=\Lambda/l^*$. 
Using the definition of the Bessel function
\eq
J_n(z)=\int\limits_{-\pi}^{\pi}{d\t\over2\pi}e^{iz\sin{\t}-in\t}
\ee
and fixing $\t_q=\pi/2$, we can perform the angular integration and
cast the direct and exchange matrix elements into the form (ignoring
the factor of $-(N_F(m_1)-N_F(m_2))$)
\eqa
&V_{d}(m_1m_2:n_1n_2)=K_p {\rhot_{n_1n_2}(Q)\rhot_{n_3n_4}(Q)\over Q}e^{\lam^* Q}\label{vd}\\
&V_{x}(m_1m_2:n_1n_2)=-K_p(-1)^{n_1+m_2}\times\nonumber\\
&\int\limits_{0}^{\infty} dSe^{-\lam^* S} \rhot_{m_1n_1}(S)\rhot_{n_2m_2}(S)J_{m_1-m_2+n_2-n_1}(QS)\label{vx}
\eea
Here $K_p={e^2\over l_0(2p+1)^{5/2}}$ is a common energy scale. It turns out that the factor $K_p$ can also be extracted from the HF energies to get
\eq
e(m)={\e(m)\over K_p}=\half\int\limits_{0}^{\infty} dS \sul_{n}|\rhot_{nm}(S)|^2(1-2N_F(n))
\ee

Since the overall factor of $K_p$ can be extracted from the entire
TDHF hamiltonian, it follows that the  ME dispersions scale
with $K_p$. Now suppose we had analyzed the fraction
$\nu(p,r)=p/(2pr+1)$. Everything above would have followed unchanged
except for the replacement $2p+1\ra 2pr+1$, and hence  $K_p$
and $\lam^*$ would change, but nothing else (it is important to note that the
ground state still consists of $p$ filled CF-LLs). Thus, in our
approach, the ME dispersions at the same $p$, but different $r$, are
related by scaling. This is an extention of the scaling relations for
gaps that we verified in an earlier publication\cite{scaling}. This
scaling holds exactly in our theory provided the constraints are
ignored. If the constraints are taken into account by cutting off the
CF-LLs, then of course a $r$ dependence will creep into the ME
dispersions.

\subsection{Spectral Function and Spectral Weight}
\medskip

It is frequently of interest to determine how much of the spectral
weight is being carried by various modes, and in particular, the
lowest mode. Such considerations are crucial for the success of the
single mode approximation (SMA)\cite{GMP}, which is known to be good near the
roton minimum for $\third$. Let us go back to the TDHF hamiltonian of
Eq.(\ref{tdhfham}), and assume that we have diagonalized it into modes
$\a$
\eqa
H^{TDHF}_{m_1m_2:n_1n_2}(\vq) \Psi^{R\a}_{n_1n_2}(\vq)&= E_{\a}(\vq)\Psi^{R\a}_{m_1m_2}(\vq)\\
\Psi^{L\a}_{n_1n_2}(\vq)H^{TDHF}_{n_1n_2:m_1m_2}(\vq)&=E_{\a}(\vq)\Psi^{L\a}_{m_1m_2}(\vq)
\eea
where the sum over ${n_1n_2}$ is implicit. Note that since $H^{TDHF}$
is not hermitian we need to consider both right and left
eigenvectors.

Now we can go back and add an external time-dependent potential
$V(\vq,\o)e^{-i\o t}$ that couples to the electron density in  the
hamiltonian. It is easy to verify that the equation for the response
of the magentoexciton operators to the external potential is
\eqa
(\o \d_{m_1m_2:n_1n_2} + &H^{TDHF}_{m_1m_2:n_1n_2})\Xb_{n_1n_2}(\vq,\o) = \nonumber\\
(N_F(m_1)-N_F(m_2))&{V(\vq,\o)\over l^{*2}L^2}\rho_{m_2m_1}(\vq)
\eea
 We can
resolve any function of $H^{TDHF}$ in terms of its eigenvectors and
eigenvalues, and in particular, we can invert the above equation to
obtain
\eqa
\Xb_{m_1m_2}(\vq,\o)=\sum\limits_{\a:n_1n_2}\Psi^{R\a}_{m_1m_2}(\vq){1\over\o+E_{\a}}\Psi^{L\a}_{n_1n_2}(\vq)&\nonumber\\
\times{(N_F(n_1)-N_F(n_2))\over l^{*2}L^2}\rho_{n_2n_1}(\vq)V(\vq,\o)&
\eea

Recalling the expansion of the density in terms of the magnetoexciton
operators, Eq.(\ref{rhob}), we obtain for the density response
\eqa
\langle \rhob \rangle=\sum\limits_{\a:m_1m_2n_1n_2} 
\rho_{m_2m_1}(\vq)\Psi^{R\a}_{m_1m_2}(\vq){1\over\o+E_{\a}}\Psi^{L\a}_{n_1n_2}(\vq)&\nonumber\\
 \times{(N_F(n_1)-N_F(n_2))\over l^{*2}L^2}\rho_{n_2n_1}(\vq)V(\vq,\o)&
\eea

Now we use the fact that due to the peculiar structure of $H^{TDHF}$
for spin-polarized MEs
$\Psi^{L\a}_{m_1m_2}=(N_F(m_1)-N_F(m_2))\Psi^{R\a}_{m_1m_2}$. Further,
using the definition of the reduced density matrix elements,
Eq. (\ref{rhot}), we obtain the susceptibility to be
\eq
\chi(\vq,\o)=\sum\limits_{\a} {A_{\a}\over \o+E_{\a}}
\ee
where
\eq
A_{\a}={\bigg(\sum\limits_{m_1m_2}\rhot_{m_2m_1}(\vq)\Psi^{R\a}_{m_1m_2}(\vq)\bigg)^2\over2\pi(2p+1)^2l^{*2}}
\ee

In the TDHF approximation all the modes are sharp, and do not
decay. The spectral function is then a sequence of delta functions at
the energies $\pm E_{\a}$ with weight $A_{\a}$. Of course, if we go
beyond TDHF, all the modes except the lowest will acquire a finite
width. Even the lowest mode may have a dispersion which kinematically
permits a quantum of high $q$ to decay into two or more quanta of
lower $q$.

To summarize, our approach to computing the spin-polarized ME
dispersions will be to compute the TDHF hamiltonian matrix in the
subspace of MEs, and to diagonalize it. If constraints are ignored,
this subspace is infinite-dimensional, since there is an infinite
tower of CF-LLs. We will follow the usual prescription of increasing
the number of CF-LLs kept till the answer converges. For small
$\lam$ we find that the gap is unphysically large, and diverges as
$\lam\ra 0$. Here we will follow the discussion presented  earlier
and keep a finite number of CF-LLs, approximately equal to $2p+1$,
to get a qualitative idea of the predictions of our theory in this
regime. It should be emphasized that we trust the results only for
$\lam\ge1.5$ or so. Fortunately, this seems to be close to   the
physical regime for high-density samples\cite{jain-th,morf-comment}.

Let us now turn to the spin-flip MEs. 
\medskip
\subsection{Spin-Flip Magnetoexcitons}
\medskip

In order to incorporate the spin quantum number we must go back to the
beginning and ask how the flux attachment is accomplished in the
initial representation. As implicit in the early wave-function
formulation of Halperin\cite{hal-spin}, and pointed out explicitly by
Lopez and Fradkin\cite{lopez2}, there are many inequivalent ways to
attach flux, where particles with like spin see each other as carrying
$2r_1$ units of statistical flux, while those with unlike spin see
each other as carrying $2r_2$ units. We will work in the picture where
the number of flux units seen by other particles is independent of the
spin, and will deal with the case of $r_1=r_2=1$ for simplicity. For
this case, the statistical gauge field remains a scalar in the spin
indices, and still satisfies
\eq
\gr\times\va(\vrr)=4\pi \rho(\vrr)
\ee
The sequence of canonical transformations\cite{us1,us2} goes through
exactly as before, and we arrive at the final representation of the
electron density projected to the low-energy subspace, expressed in
second quantized form as
\eq
{\hat\rhob}(\vq)=\sum\limits_{\s,n_1n_2} \rho_{n_2n_1}(\vq)\hX_{\s,n_1n_2}(\vq)
\ee
where we define the spin-$\s$ ME operators 
\eq
\hX_{\s,n_1n_2}(\vq)=\sum\limits_{k_1k_2} (x_{\nu_2\nu_1}(\vq))^* d^{\dagger}_{\s,\nu_1}d_{\s,\nu_2}
\label{meopspol}\ee
In other words the total density is the sum of the $\ua$ and
$\da$ charge densities. One can also define the spin-density
operators projected to the low-energy subspace 
\eqa
&\hS_z(\vq)=\sum\limits_{\s,k_1k_2} (x_{\nu_2\nu_1}(\vq))^* \rho_{n_2n_1}(\vq){\s\over2}d^{\dagger}_{\s,\nu_1}d_{\s,\nu_2}\\
&\hS_+(\vq)=\sum\limits_{k_1k_2} (x_{\nu_2\nu_1}(\vq))^* \rho_{n_2n_1}(\vq)d^{\dagger}_{\ua,\nu_1}d_{\da,\nu_2}\\
&\hS_-(\vq)=\sum\limits_{k_1k_2} (x_{\nu_2\nu_1}(\vq))^* \rho_{n_2n_1}(\vq)d^{\dagger}_{\da,\nu_1}d_{\ua,\nu_2}
\eea
The projected spin and charge operators no
longer commute with each other, unlike the full spin and charge
operators. Instead they form a representation (to lowest leading order
in $q$) of the LLL spin-charge algebra discussed previously in
ref.\cite{moon}. 
For our purposes, it is useful to define the spin-flip 
ME operators 
\eqa
&\hY_{+,n_1n_2}(\vq)=\sum\limits_{k_1k_2} (x_{\nu_2\nu_1}(\vq))^* d^{\dagger}_{\ua,\nu_1}d_{\da,\nu_2}\\
&\hY_{-,n_1n_2}(\vq)=\sum\limits_{k_1k_2} (x_{\nu_2\nu_1}(\vq))^* d^{\dagger}_{\da,\nu_1}d_{\ua,\nu_2}
\eea

We will be dealing exclusively with fully polarized states for
simplicity, though the method can be trivially extended to partially
polarized states as well. Let us arbitrarily designate $\ua$ as the
sign of the majority spin. There will be an additional Zeeman energy
of $E_Z(n_{\da}-n_{\ua})/2$. Since the hamiltonian commutes with the
 $z$-component of the total spin, this energy can be added to the
spin-flip excitations at the end. We will therefore ignore the Zeeman
term when computing the dynamics. As in the previous subsection, we
commute the hamiltonian with $\hY_{\pm}$ and find the hamiltonian
matrix in the subspace of spin-flip MEs. The final result is

\eqa
&[\hH,\hY_{-,m_1m_2}(\vq)]=(\e_{\da}(m_1)-\e_{\ua}(m_2))\hY_{-,m_1m_2}(\vq)\nonumber\\
&-(N_{F\da}(m_1)-N_{F\ua}(m_2))\sul_{n_1n_2}\hY_{-,n_1n_2}(\vq)V_{x}(m_1m_2;n_1n_2)\nonumber\\
&
\eea
where $V_{x}$ is the same exchange matrix element as in
Eq. (\ref{vx}). We also define the spin-dependent HF energies as

\eq
\e_{\s}(m)={\pi\over L^2}\intq v(q)\sum\limits_{n}(1-2N_{F\s}(n))|\rho_{nm}(\vq)|^2
\ee
where $N_{F\s}(m)$ denotes the spin-dependent Fermi occupation of the
single-particle state $m$.  Note that only modes which have
$(N_{F\da}(m_1)-N_{F\ua}(m_2))\ne 0$ need be considered. In the fully
polarized states $N_{F\da}(m)=0$ for all $m$, and this restricts us to
consider $N_{F\ua}(m)=1$. The ``negative energy'' MEs appearing in the
previous subsection are absent here, and the TDHF hamiltonian matrix
is hermitian. 

Let us briefly note that the formalism developed in this section can
be directly applied to the very interesting case of double-layer
quantum Hall systems, which have been the subject of much recent
interest\cite{double-layer-th,double-layer-ex}.

We are now ready to present the results.

\bigskip
\section{Results}
\label{results}
\bigskip

\subsection{Gaps}
\medskip

In a previous publication Shankar and the present author have
presented results for the transport gap in a few
fractions\cite{single-part}, calculated using the methods of section
II. We pause here to view the gaps in a slightly different light. In
many field-theoretic approaches to computing the ME dispersions it is
assumed that the interactions produce CF-LLs whose energy dependence
on the CF-LL index $n$ is cyclotron-like
\eq
E_n=n\o_c^{*}\label{cychyp}
\ee
Indeed, in the CS approach of Lopez and Fradkin\cite{lopez}, this is
the exact result in lowest order, with the bare mass $m$ setting the
scale for $\o_c^*$.  This seems to be a natural assumption.  On the
other hand, in the LLL, all splittings are the result of interactions,
and there is no {\it a priori} reason why they should follow
Eq. (\ref{cychyp}). In our approach we can compute these energies, and
can therefore test this hypothesis. In Table 1 we present the energies
of the CF-LLs for $\nu=\third$, for $n_{max}=5$. We have set the
energy of the $n=0$ CF-LL to 0, and normalized all energies in units
of $E_{n=1}$. The cyclotronic prediction for this normalization is
$E_n=n$.

\vskip 0.1in
\begin{tabular}{|c|c|c|c|c|c|}
\hline
$\lam$ & $E_1$ & $E_2$ & $E_3$ & $E_4$ & $E_5$ \\
\hline
0.0 & 1.00 & 1.75 & 2.37 & 2.88 & 3.32 \\
\hline
1.0 & 1.00 & 1.48 & 1.76 & 1.90 & 1.89 \\
\hline
2.0 & 1.00 & 1.32 & 1.50 & 1.54 & 1.43 \\
\hline
3.0 & 1.00 & 1.22 & 1.35 & 1.38 & 1.27 \\
\hline
\end{tabular}
\vskip 0.1in

It is quite clear that there are strong deviations from the
cyclotronic form. These deviations become greater as the sample
thickness increases. In particular,  for $\lam\ge1.0$ we find the
strange result that $E_5$ is less than $E_4$. This is another
manifestation of the unphysical nature of the higher CF-LLs. For
higher $\lam$ there is a distinct tendency for the energies to fall
towards $E_1$, to form almost a continuum above $E_1$. All these
features are present for the other fractions as well. 

To emphasize the collective nature of the energies, we present the
spin-dependent HF energies in the absence of Zeeman interactions for
$\nu=\twof$ at $\lam=1.0$, with $n_{max}=4$ in Table II.

\vskip 0.1in
\begin{tabular}{|c|c|c|c|c|c|}
\hline
$n$ & 0 & 1 & 2 & 3 & 4 \\
\hline
$E_{n\ua}/K_2$ & -0.36 & 0.81 & 2.62 & 3.39 & 3.40 \\
\hline
$E_{n\da}/K_2$ & 2.35 & 3.74 & 4.68 & 5.20 & 5.07 \\
\hline
\end{tabular}
\vskip 0.1in

Here $K_2$ is an example of the constant $K_p$ defined in Section
IIIB. The majority ($\ua$-spin) and minority ($\da$-spin) energies are
quite different for the same CF-LL index, even when the Zeeman
coupling is zero, because the majority spin energy has a large
negative exchange contribution, while the minority spin energy has
only a Hartree contribution. Neither of the energies has a cyclotronic
form.

We therefore come to the conclusion that the cyclotronic form of the
energy should be regarded with skepticism.

\subsection{Spin-Polarized Magnetoexcitons}
\medskip

\begin{figure}
\narrowtext
\epsfxsize=2.4in\epsfysize=2.4in
\hskip 0.3in\epsfbox{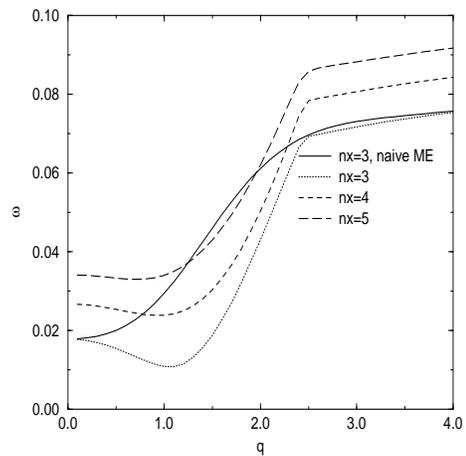}
\vskip 0.15in
\caption{Spin-polarized ME dispersions for $\nu=1/3$ at $\lam=1.5$ 
for different values of $n_{max}$.  All energies are in units of
$e^2/\varepsilon l$. The solid line is the energy of the ``naive'' ME, which
corresponds to the strong-field limit in the IQHE. 
\label{fig1}}
\end{figure}

Figure \ref{fig1} shows the ME dispersions for $\nu=\third$, at a
$\lam=\Lambda/l_0=1.5$. We are using a cutoff on the number of
CF-LL's, $n_{max}$, denoted by ``nx'' in the figure. The solid line
denotes the ``naive'' ME. In other words, we take the state created by
the ME operator $\hX_{10}(\vq)$ and just compute its energy as a
function of $q$. The energy of the naive ME corresponds to the
strong-field limit for the IQHE.  It is clear that including the other
ME's, and the interaction between them, drastically alters the
dispersion. In particular, the magnetoroton minimum present in the
other dispersions is missing in the naive one. We conclude that
the strong-field limit is not justified in the FQHE.

Let us now turn to the evolution of the dispersions with $q$ and
$n_{max}$. As $q$ increases the energy asymptotically tends to the
transport gap, which itself increases with $n_{max}$. The gap at
$n_{max}=5$ is about half its value for $n_{max}=\infty$. However, the
gap at $n_{max}=5$ is about 20\% {\it greater } than the true
gap\cite{jain-th} for $\lam=1.5$. 
\begin{figure}
\narrowtext
\epsfxsize=2.4in\epsfysize=2.4in
\hskip 0.3in\epsfbox{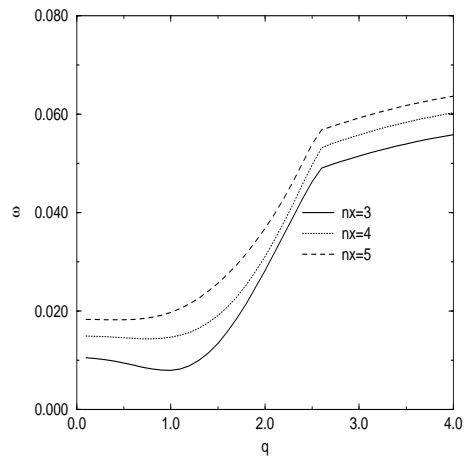}
\vskip 0.15in
\caption{Spin-polarized ME dispersions for $\nu=1/3$ at $\lam=2.0$ for
different values of $n_{max}$.
\label{fig2}}
\end{figure}
The roton minimum is very shallow
and the value of the energy at $q=0$ is quite small. The $q$ at the
minimum is about $1.1l_0^{-1}$, slightly smaller than the
corresponding value for the pure Coulomb interaction. This is
qualitatively consistent with the results of the
SMA\cite{GMP}. However, for $q=0$ the quasiparticle and the quasihole
are right on top of each other, and our charge density operator may
not have enough short distance structure to get the right numbers for
small $\lambda$. For large $\lam$, however, the potential becomes very
flat at short distances, and we expect better results.

Fig. \ref{fig2} shows the same picture for $\lam=2.0$. As can be seen,
the fractional change in the transport gap as $n_{max}$ increases is
even smaller here (the $n_{max}=\infty$ value of the gap is now about
50\% higher than that for $n_{max}=5$, which is itself 20\% higher than the
true gap\cite{jain-th}). The magnetoroton minimum is even smaller and
seems to shift slightly towards smaller $q$.

\begin{figure}
\narrowtext
\epsfxsize=2.4in\epsfysize=2.4in
\hskip 0.3in\epsfbox{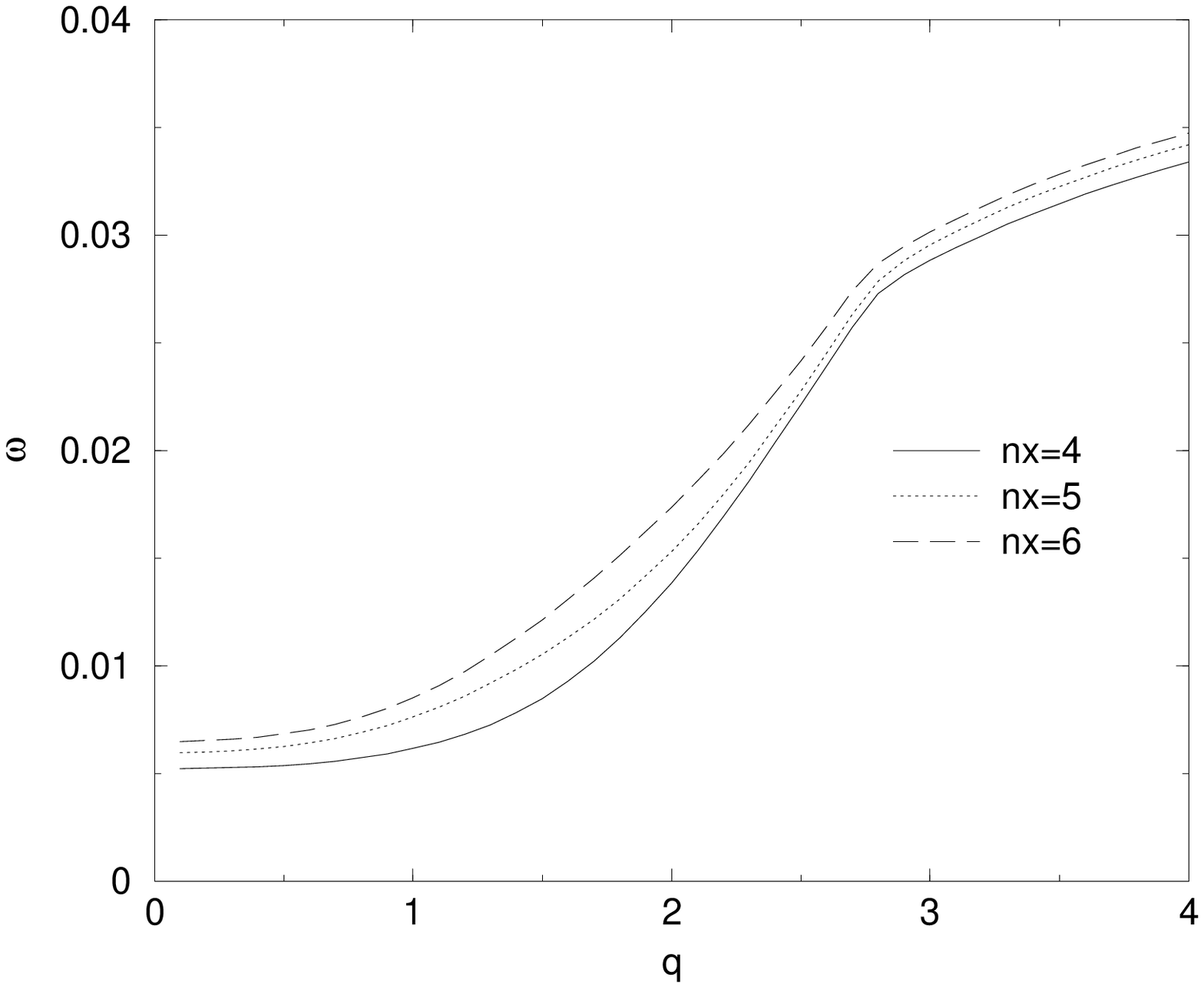}
\vskip 0.15in
\caption{Spin-polarized ME dispersions for $\nu=1/3$ at $\lam=3.0$ for
different values of $_{max}$.
\label{fig3}}
\end{figure}

To continue the trend towards greater thickness we show the
corresponding results for $\lam=3.0$ in Fig. \ref{fig3}, for which
the $n_{max}=\infty$ value of the gap is is 35\% higher than that for
$n_{max}=5$, which is itself within 10\% of the true gap. The minimum
at finite $ql$ has disappeared and the minimum is now at $q=0$.

\begin{figure}
\narrowtext
\epsfxsize=2.4in\epsfysize=2.4in
\hskip 0.3in\epsfbox{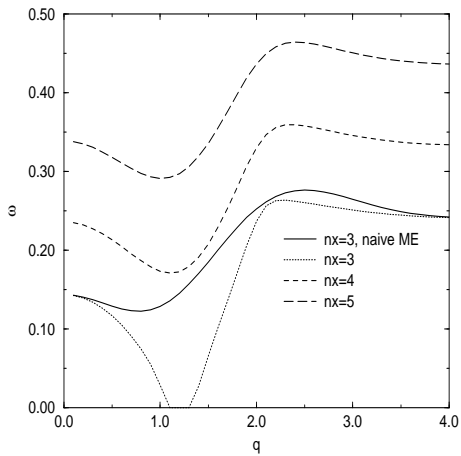}
\vskip 0.15in
\caption{Spin-Polarized ME dispersions for $\nu=1/3$ at $\lam=0.0$ 
for different values of $n_{max}$. Note that the dispersion shows an
instability for $n_{max}=3$, but none for $n_{max}>3$.
\label{fig4}}
\end{figure}

Fig. \ref{fig4} shows the same calculation done for the pure Coulomb
interaction, $\lam=0.0$. For $n_{max}=3$ one finds that the dispersion
becomes unstable for a range of $q$ between $1.1l_0^{-1}$ to
$1.35l_0^{-1}$. However, the instability disappears if $n_{max}$ is
increased, and now there is a deep magnetoroton minimum around
$ql_0=1.25$, which is the correct
position\cite{hal-rez,GMP,exact-ex,jain-ex}. However, the values of
the energy are much larger than the true values. Thus, while a
qualitatively correct picture is obtained for $n_{max}\ge 4$, we
cannot trust the numbers for the Coulomb interaction. The main reason
is probably because, as pointed out above, our $\rhob$ misses the
short-distance physics which is important for the Coulomb interaction.

\begin{figure}
\narrowtext
\epsfxsize=2.4in\epsfysize=2.4in
\hskip 0.3in\epsfbox{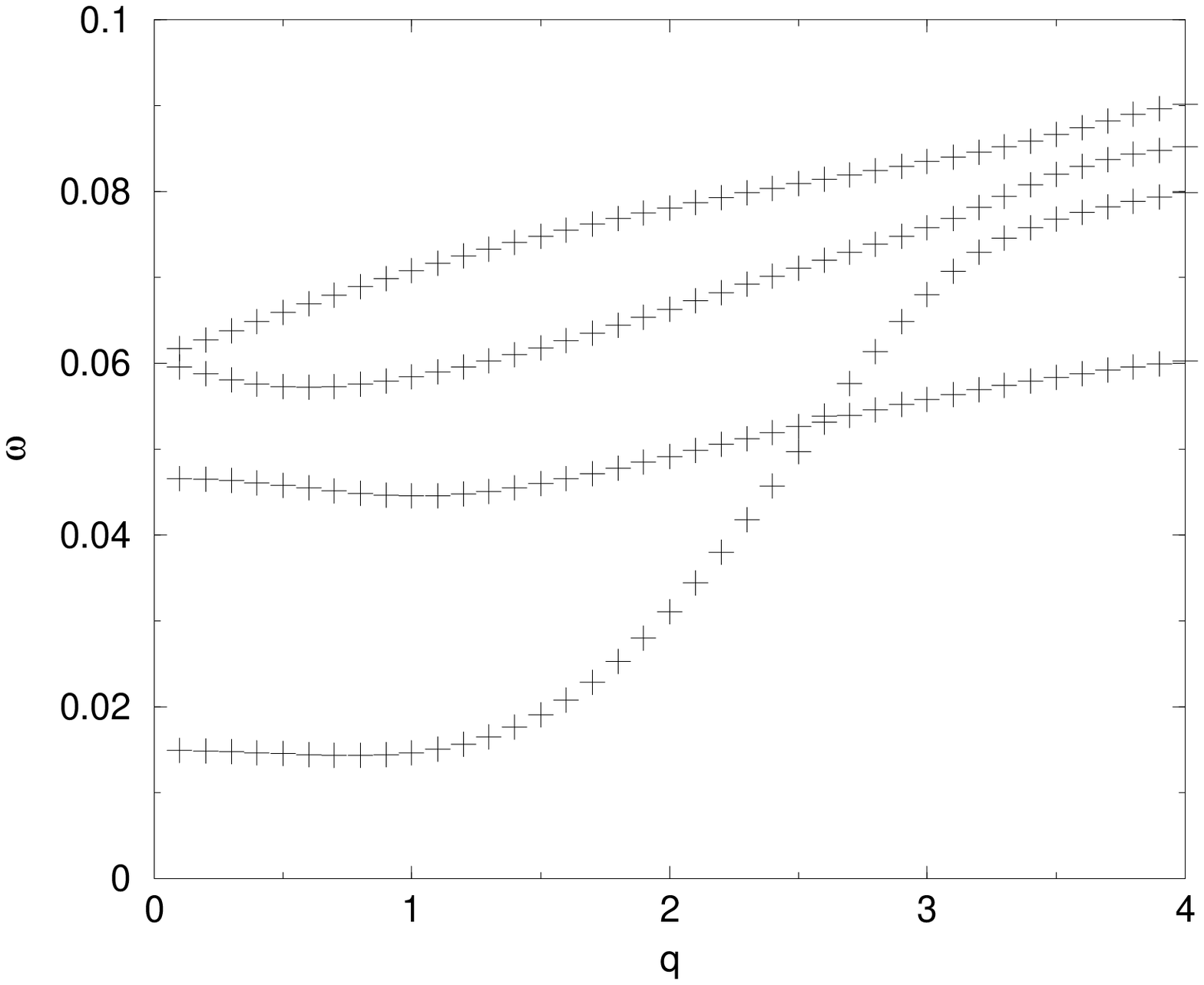}
\vskip 0.15in
\caption{All the ME dispersions for $\nu=1/3$ at $\lam=2.0$ for 
$n_{max}=4$. The lowest mode is well-separated but the others form a
quasi-continuum.
\label{fig5}}
\end{figure}

Since our approach is a dynamical theory of CFs, we can obtain all the
ME dispersions for a given $n_{max}$. To emphasize this,
Fig. \ref{fig5} shows all the ME dispersions for $\nu=\third$,
$\lam=2.0$, and $n_{max}=4$.

\begin{figure}
\narrowtext
\epsfxsize=2.4in\epsfysize=2.4in
\hskip 0.3in\epsfbox{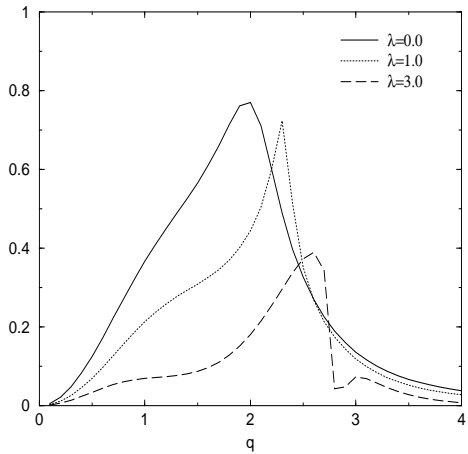}
\vskip 0.15in
\caption{Fraction of spectral weight carried by the lowest mode 
for $\nu=1/3$ at $n_{max}=4$ for different values of $\lam$. Note that
the lowest mode carries vanishing spectral weight for large and small
$q$, and that its spectral weight decreases drastically with
increasing $\lam$.
\label{fig6}}
\end{figure}

In Fig. \ref{fig6} we show the fraction of total spectral weight in
the lowest mode for different values of $\lam$. The value of $n_{max}$
has been set at 4 for all the curves. Two features stand out: Firstly,
the lowest mode carries vanishing spectral weight at small and large
$q$, regardless of $\lam$. This means that the SMA will overestimate
the gap for $q=0$ and for large $q$. Secondly, it is clear that for
small $\lam$, much of the spectral weight is in the lowest mode for
intermediate values of $ql$. These two facts are known for $\lam=0.0$
from exact diagonalizations\cite{haldane-book}. Our calculation
actually underestimates the weight for $\lam=0.0$ (which, to the
author's knowledge, is the only case where it has been previously
computed). However, our results also show that the lowest mode carries
a decreasing fraction of the spectral weight with increasing $\lam$,
which means that the SMA will become less useful for larger $\lam$.

\begin{figure}
\narrowtext
\epsfxsize=2.4in\epsfysize=2.4in
\hskip 0.3in\epsfbox{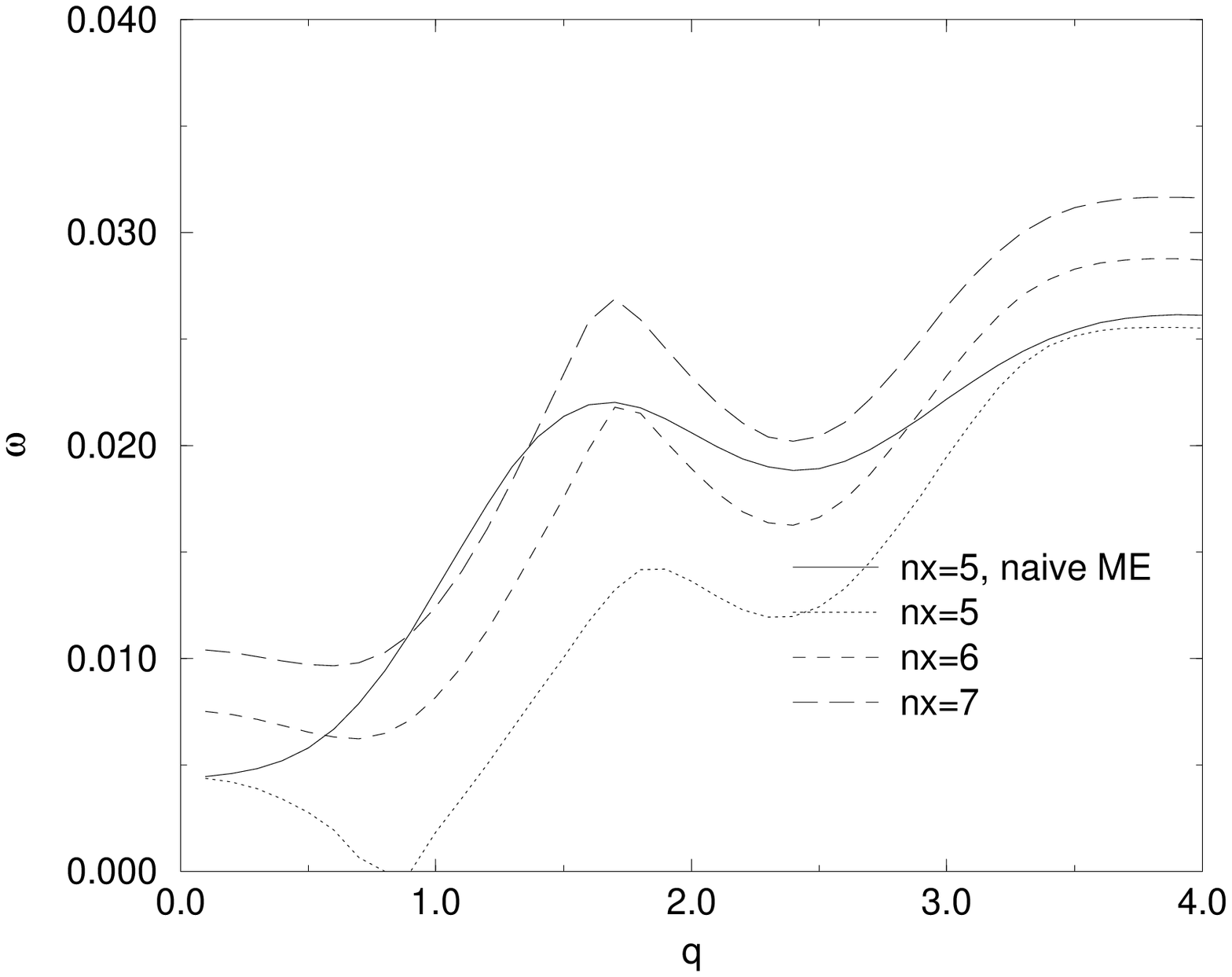}
\vskip 0.15in
\caption{Spin-polarized ME dispersions for $\nu=\twof$ at $\lam=1.5$ 
for different values of $n_{max}$.  
\label{fig7}}
\end{figure}

The next set of five figures presents results of similar calculations
for $\nu=\twof$. Here $n_{max}=5$ shows instabilities for $\lam=1.5, 2.0$,
while $n_{max}\ge 6$ shows the right behavior (we will comment in detail on
instabilities in the next subsection). The minima appear at
$ql_0=0.85, 2.4$. 

\begin{figure}
\narrowtext
\epsfxsize=2.4in\epsfysize=2.4in
\hskip 0.3in\epsfbox{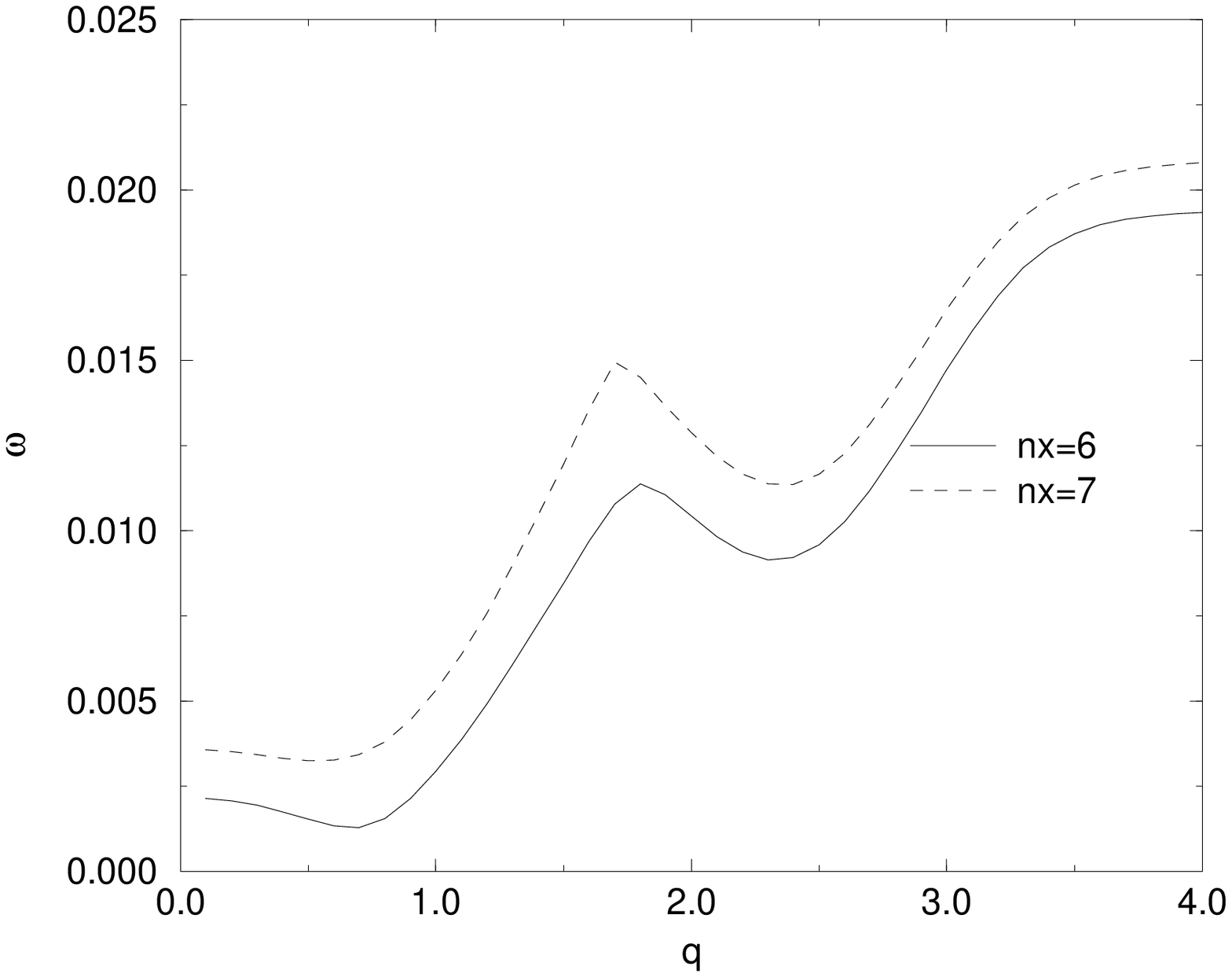}
\vskip 0.15in
\caption{Spin-polarized ME dispersions for $\nu=\twof$ at $\lam=2.0$ 
for different values of $n_{max}$.  
\label{fig8}}
\end{figure}

While the position of the first minimum is
consistent with the results of exact diagonalizations, the second
minimum seems to be too high in $q$ (exact
diagonalizations\cite{exact-ex} and other\cite{jain-ex,rpa-ex} numerical
results indicate the second minimum at $ql_0=1.6$).

\begin{figure}
\narrowtext
\epsfxsize=2.4in\epsfysize=2.4in
\hskip 0.3in\epsfbox{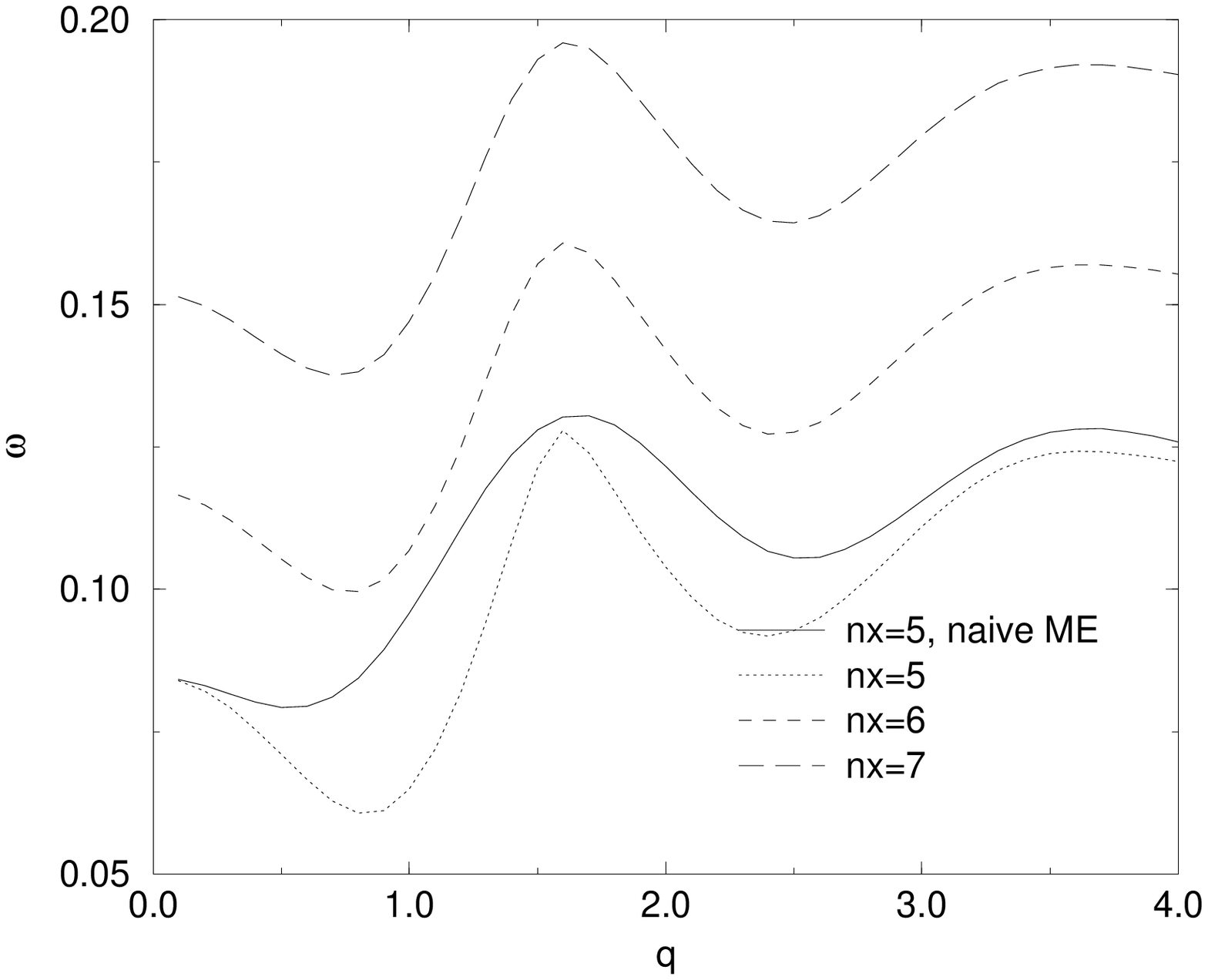}
\vskip 0.15in
\caption{Spin-polarized ME dispersions for $\nu=\twof$ at $\lam=0.0$ 
for different values of $n_{max}$.  
\label{fig9}}
\end{figure}

\begin{figure}
\narrowtext
\epsfxsize=2.4in\epsfysize=2.4in
\hskip 0.3in\epsfbox{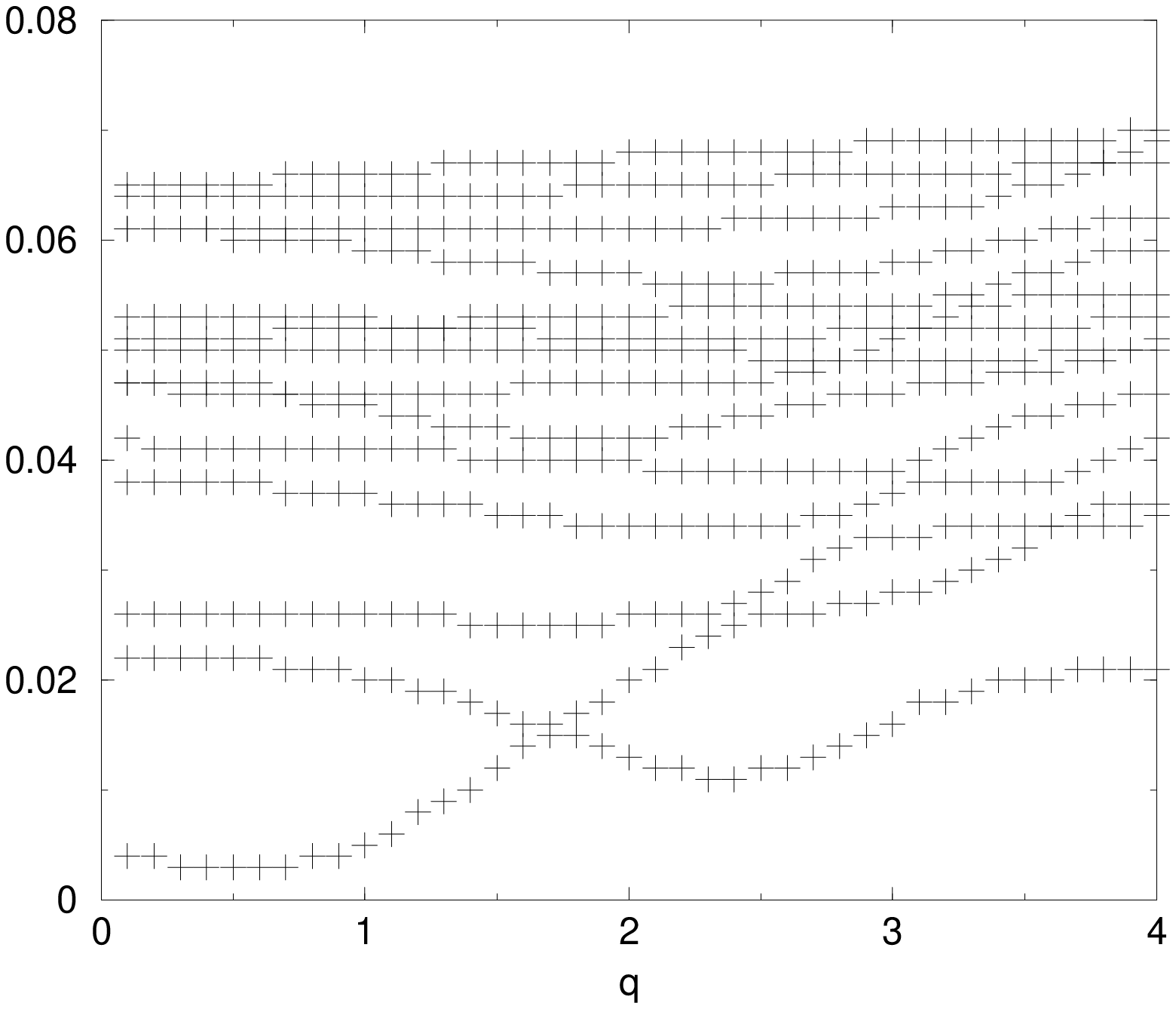}
\vskip 0.15in
\caption{All the ME dispersions for $\nu=\twof$ at $\lam=2.0$ for 
$n_{max}=8$. The lowest mode is well-separated but the others form a
quasi-continuum.
\label{fig10}}
\end{figure}

Fig. (\ref{fig10}) shows all the dispersions for $\lam=2.0$ and
$n_{max}=8$. It is clear that beyond the lowest ME mode, there is
almost a continuum of excitations (recall that these are by no means
all the excitations; these are only the single quasiparticle-single
quasihole excitations). 

\begin{figure}
\narrowtext
\epsfxsize=2.4in\epsfysize=2.4in
\hskip 0.3in\epsfbox{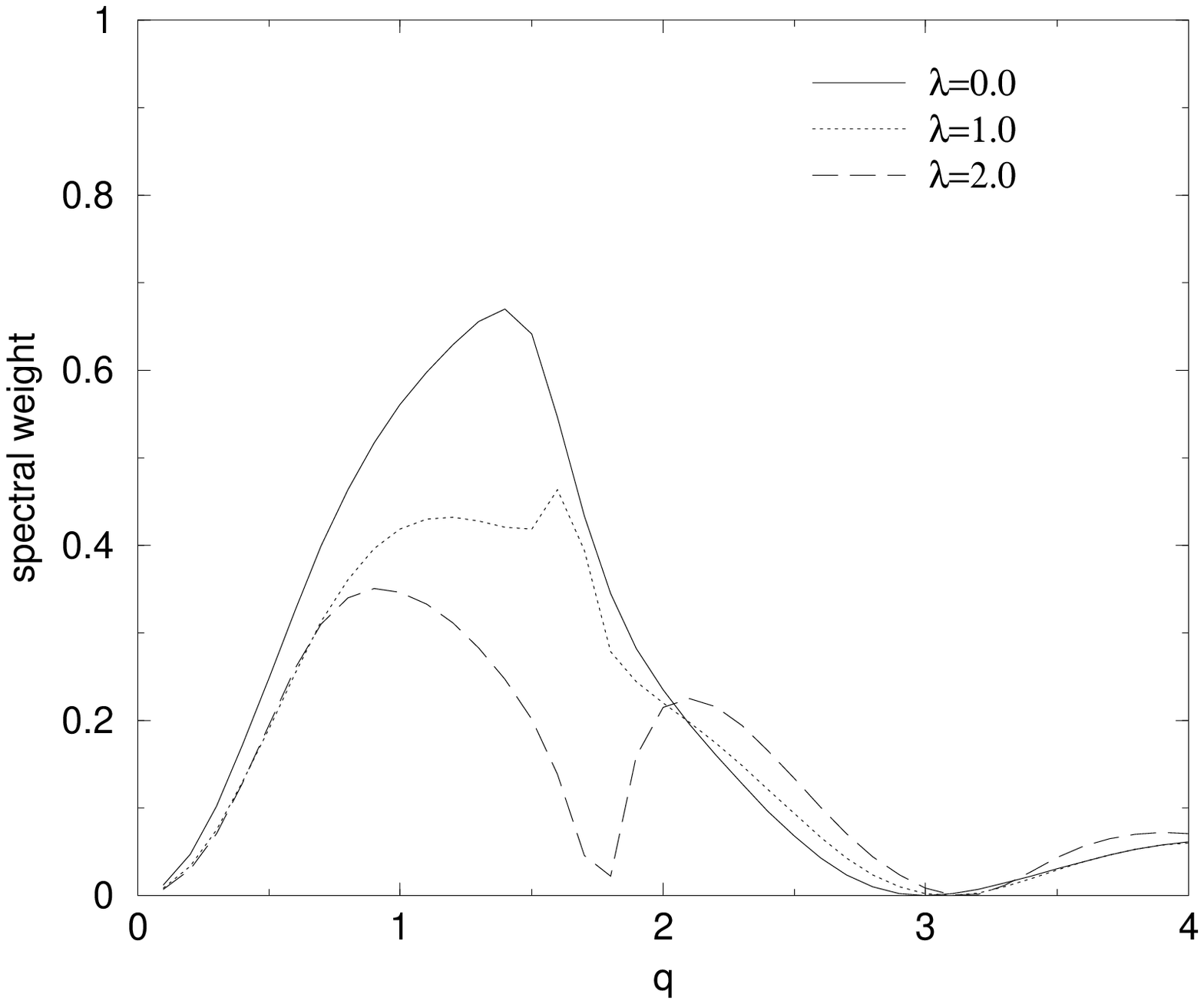}
\vskip 0.15in
\caption{Fraction of spectral weight carried by the lowest mode 
for $\nu=\twof$ at $n_{max}=7$ for different values of $\lam$. Just as
in the $\nu=\third$ case the lowest mode carries vanishing spectral
weight for large and small $q$, and its spectral weight decreases with
increasing $\lam$.
\label{fig11}}
\end{figure}

In Fig. \ref{fig11} we present the fraction of the spectral weight in
the lowest mode for $n_{max}=7$. It is clear that this is lower for
all $q$ at $\lam=0.0$ than in the case of $\third$. Once again, as
$\lam$ increases the fraction of the spectral weight in the lowest
mode decreases.

Finally, we turn to $\nu=\threes$, for which we present results for
$n_{max}=10$ only, and for values of $\lam=0.0, 1.0, 2.0$.  We find
that the position of first minimum agrees with that from exact
diagonalizations\cite{exact-ex} and CF-wavefunction
calculations\cite{jain-ex} ($ql\approx0.6$), while the second and third
minima seem to be displaced to higher $q$, just as in the $\nu=\twof$
case. 

\begin{figure}
\narrowtext
\epsfxsize=2.4in\epsfysize=2.4in
\hskip 0.3in\epsfbox{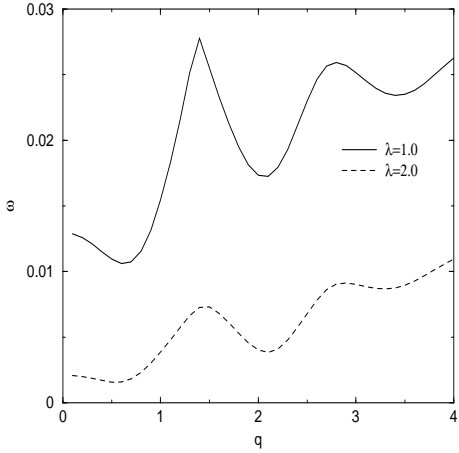}
\vskip 0.15in
\caption{Spin-polarized ME dispersions for $\nu=\threes$ at $\lam=1.0$ 
and $2.0$ for  $n_{max}=10$.
\label{fig12}}
\end{figure}

Fig. \ref{fig14} shows the spectral weight in the lowest mode,
which looks very similar to the $\twof$ case. We do not really trust
our results quantitatively for small $\lam$. Therefore, while one can
definitely say that for all fractions the spectral weight decreases as
thickness increases, it is difficult to see any other clear trend
(such as a trend with increasing denominator).

\begin{figure}
\narrowtext
\epsfxsize=2.4in\epsfysize=2.4in
\hskip 0.3in\epsfbox{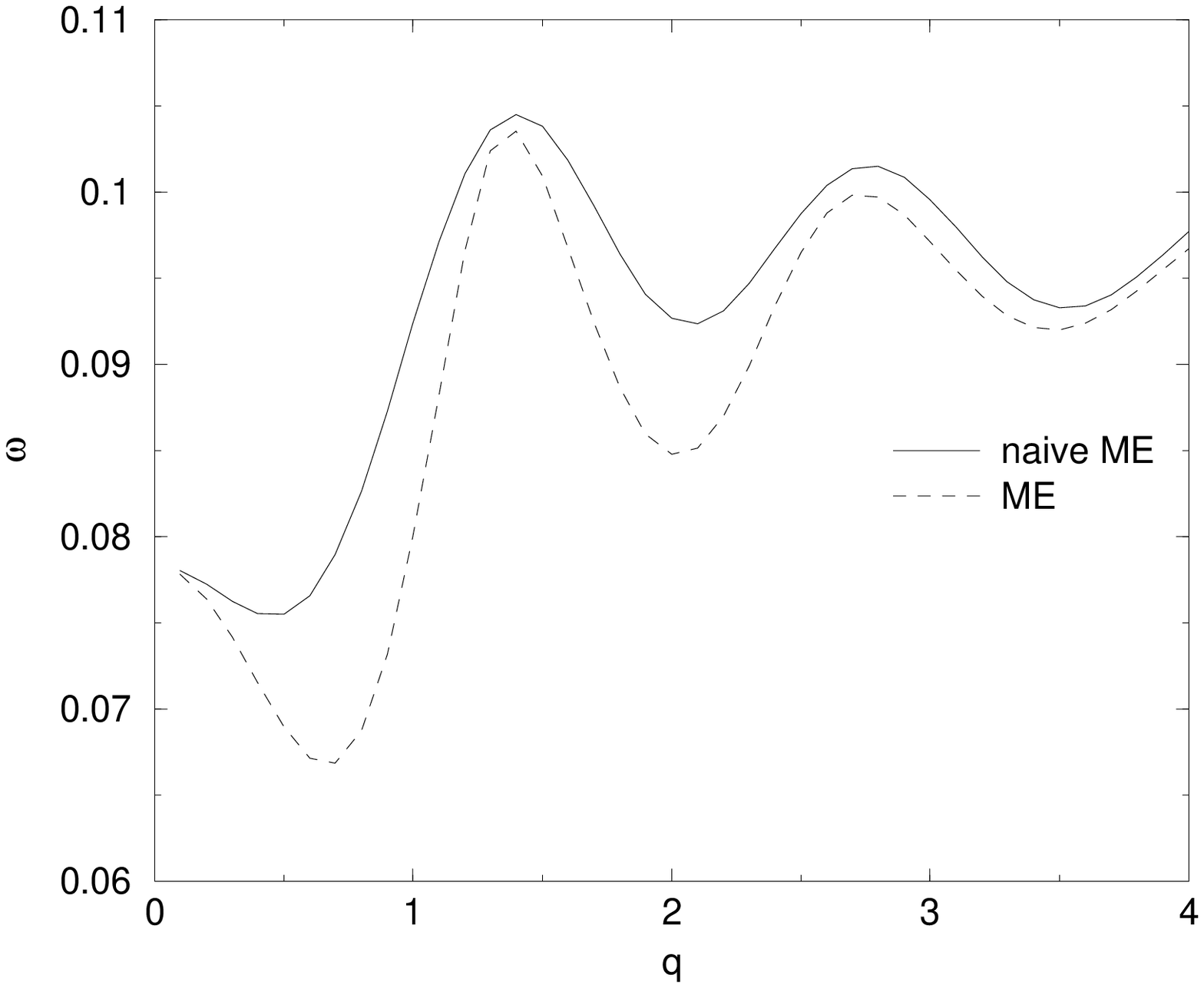}
\vskip 0.15in
\caption{Spin-polarized ME dispersions for $\nu=\threes$ at $\lam=0.0$ 
for  $n_{max}=10$.  
\label{fig13}}
\end{figure}

\begin{figure}
\narrowtext
\epsfxsize=2.4in\epsfysize=2.4in
\hskip 0.3in\epsfbox{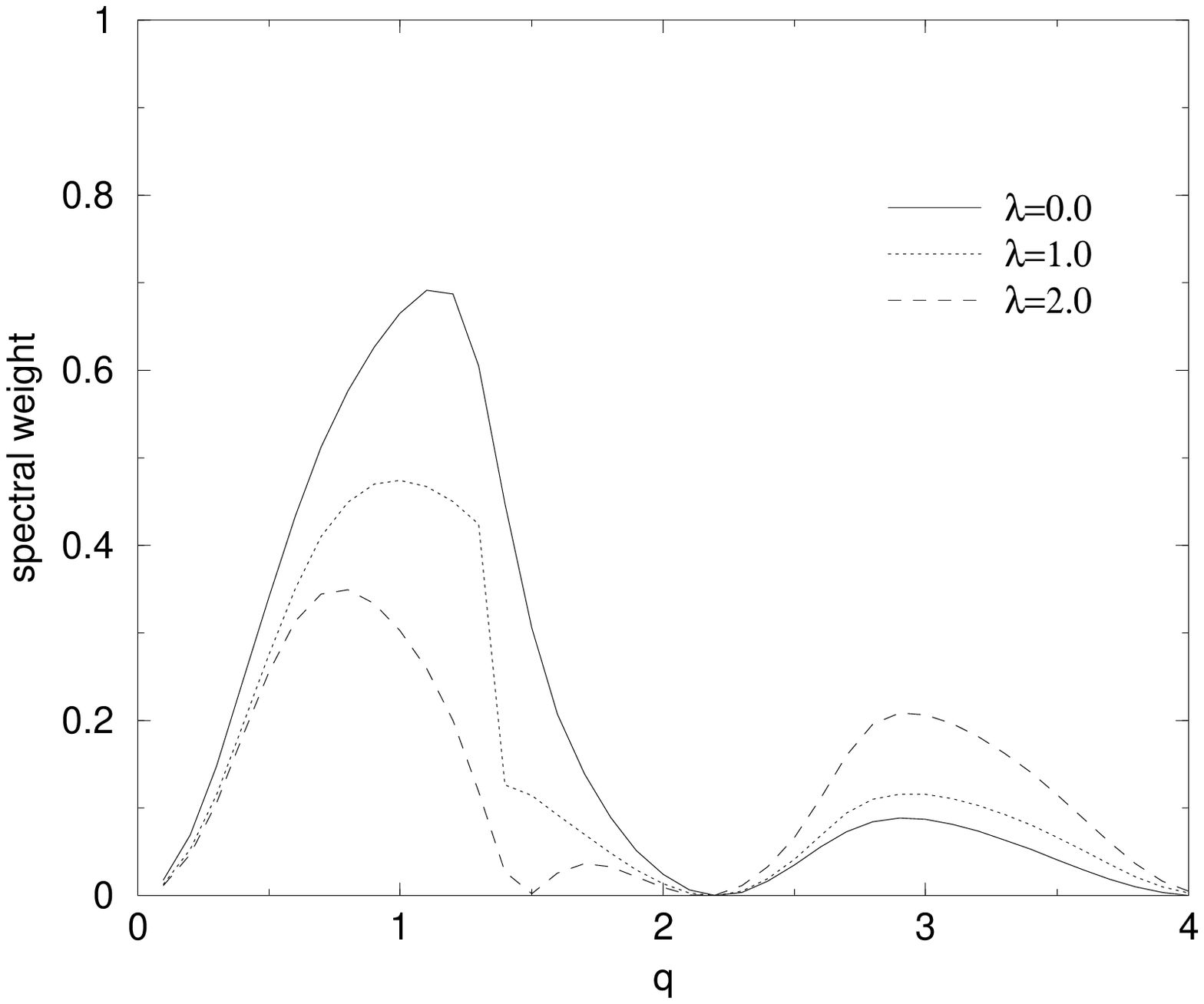}
\vskip 0.15in
\caption{Fraction of spectral weight carried by the lowest mode 
for $\nu=\threes$ at $n_{max}=10$ for different values of $\lam$. Just as
in the $\nu=\third$ case the lowest mode carries vanishing spectral
weight for large and small $q$, and its spectral weight decreases with
increasing $\lam$.
\label{fig14}}
\end{figure}

In conclusion, our results with a cutoff on $n_{max}$ appear to be in
qualitative agreement with numerical results for $\lam=0$. However,
there are quantitative discrepancies which make them unreliable for
small $\lam$. The reasons for the failure of our theory to reproduce
numbers at small $\lam$ are well-understood, and have been repeatedly
mentioned above.

On the other hand, we expect our theory to be quantitatively good at
large $\lam$. Recall that for $2.0\le\lam\le3.0$ our
results\cite{single-part} for the transport gap were within 30\% of
the numerical values\cite{jain-th}. What needs to be done is to
compare our results to numerical ones at large $\lam$. Unfortunately,
exact diagonalizations and other numerical methods have so far
concentrated on the pure Coulomb interaction, and few results on ME
dispersions are available for large $\lam$. In the absence of exact
results, our results are predictions for the ME dispersions.

We now turn to a brief discussion of instabilities.

\subsection{Instabilities Inferred from Magnetoexciton Dispersions}
\medskip

The use of ME dispersions to study instabilities was pioneered by
Girvin, MacDonald, and Platzman in their seminal paper on the
SMA\cite{GMP}. While they did not find actual instances where the ME
dispersion became negative, they did find that for $\nu=1/7$, the
magnetoroton minimum nearly touched zero. On this basis, they
conjectured that in the presence of disorder, the gaps would be
reduced, and since the potential instability is at finite wavevector,
that there would be a quantum phase transition to a Wigner
crystal\cite{wc-instability}. Kamilla and Jain\cite{jain-wc} have
demonstrated within the CF-wavefunction approach that for $\nu=1/9$
for the pure Coulomb interaction, the ME dispersion indeed becomes
negative at a finite wavevector, signalling an instability of the
incompressible liquid. The transition is expected to be first order,
and probably occurs before ({\it i.e.} for $\nu$ greater than the
$\nu$ for which) the ME dispersion becomes negative.

Our interest will be in instabilities as a function of the thickness
parameter $\lam$. While one can always make a CF variable by fiat, by
attaching two statistical flux quanta to an electron, the experimental
reality of the CF depends on energetics. It is helpful to think of the
case of $\nu=\third$ in terms of the Haldane pseudopotentials
$V_m$\cite{haldane-book}. These pseudopotentials give the energy of a
pair of electrons in the LLL with relative angular momentum $m$. The
Laughlin state has no pairs in a relative $m=1$ state. The reason this
is such a good trial state for the Coulomb potential is that $V_1$ is
much bigger than $V_3$, and there are no other states without pairs in
$m=1$. This is also why the state is incompressible: Every excitation
must necessarily have an amplitude for creating a pair in an $m=1$
state, and costs a large energy of order $V_1$. Now let us make the
sample thicker. As $\lam$ increases the ratio $V_3/V_1$ increases, and
at some critical value of this ratio, the system will make a
transition to a compressible state\cite{hal-rez}. This transition is
expected to be first order\cite{hal-rez}, but one may still look for
signals of this transition in the incompressible state.

In this spirit, I have analyzed the ME dispersions as a function of
both $\lam$ and $n_{max}$. First consider the $\nu=1/3$ state. I find
that while there appear to be instabilities at finite $q$ for small
$n_{max}$ (an example is Fig. \ref{fig4}), they disappear when one
increases $n_{max}$. So I find no reliable signs of instability for
$\nu=\third$. However, for $\nu=\twof$, I find that at $\lam_{cr}=2.65$ the ME
dispersion becomes zero at $ql\approx0.4$. I have increased $n_{max}$
to $10$ (much larger than the rough cutoff of $2p+1=5$ from
section{}), and observed that the instability persists. Similarly, I
find that for $\nu=\threes$, there is a finite wavevector instability
at $\lam_{cr}=2.4$. It is interesting that the $\lam_{cr}$ is not the
same for for fractions which share the same number of flux attached,
but decreases with increasing denominator.

Experimentally, such instabilities would be seen as the lack of a
fractionally quantized plateau beyond a certain sample density, even
at zero temperature. Presumably, the above numbers for $\lam_{cr}$ are
overestimates for the experimental values of the thickness parameter,
since disorder, which has not been taken into account here, will
further reduce the gaps.

Let us now turn to the spin-flip ME dispersions. 

\subsection{Spin-Flip Magnetoexcitons}
\medskip

\begin{figure}
\narrowtext
\epsfxsize=2.4in\epsfysize=2.4in
\hskip 0.3in\epsfbox{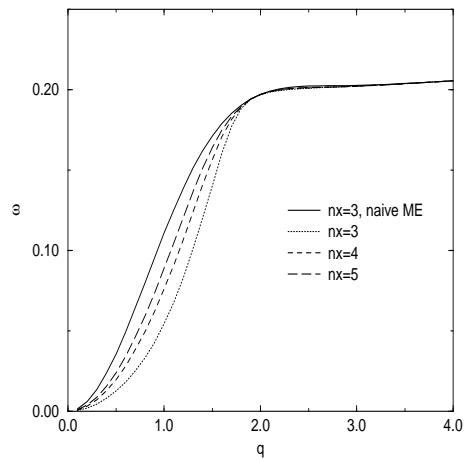}
\vskip 0.15in
\caption{Spin-wave dispersion for $\nu=\third$ at $\lam=0.0$ for 
different values of $n_{max}$. As $n_{max}$ increases the separation
between levels increases, decreasing the interaction between naive
MEs. 
\label{fig15}}
\end{figure}

Figure \ref{fig15} shows our results for $\nu=\third$ for the pure
Coulomb interaction in the absence of a Zeeman term. What is shown is
the lowest branch, which corresponds in this case to the spin-wave
mode. In accordance with Goldstone's theorem, the energy vanishes as
$q\to 0$ (quadratically, as it should for a ferromagnet). Upto about
$ql_0\approx1$ the dispersion for $n_{max}=3$ agrees quite well with
the exact diagonalization results of Nakajima and Aoki\cite{nakajima},
and thus also gives approximately the correct spin stiffness.

Once again, the
``naive'' spin-flip ME is higher in energy than the true mode by
almost a factor of two in this region, and interaction between the
various naive modes results in closer agreement with exact results. In
this context the accuracy of the results of Nakajima and
Aoki\cite{nakajima} based on flux attachment appear puzzling, since
the strong-field limit is implicit in their approach. On the other
hand, $\lam=0.0$ is the worst case for our theory, and perhaps it is
our agreement with the spin-wave energy that is more surprising. It is
difficult to compare our results to Mandal's work\cite{mandal},
since in the spirit of earlier work in the fermionic CS
approach\cite{lopez,rpa-ex} a phenomenological effective mass
appears, and the cyclotronic hypothesis is used for the energies. 

As $q\to\infty$ the ME energy tends asymptotically to the
spin-reversed gap (which is considerably overestimated by our theory
at $\lam=0.0$). It is important to note that all higher spin-flip
modes are well separated from the lowest one for small $q$, which
justifies identifying the low-energy dynamics of the polarized
$\third$ state with that of the quantum ferromagnet\cite{q-fm-th}.

\begin{figure}
\narrowtext
\epsfxsize=2.4in\epsfysize=2.4in
\hskip 0.3in\epsfbox{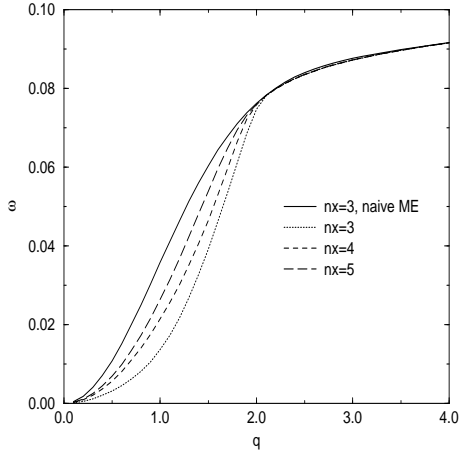}
\vskip 0.15in
\caption{Spin-wave dispersion for $\nu=\third$ at $\lam=1.0$ for
different values of $n_{max}$. 
\label{fig16}}
\end{figure}

We follow the progression as $\lam $ increases in
Fig. \ref{fig16}. Once again, it is unfortunate that numerical results
are not available for large $\lam$, where we expect our theory to be
quantitatively accurate. Note, that for the spin-reversed gap, our
prediction is about 30\% higher the exact answer\cite{bonesteel} for
$\lam$ as low as $1.5$.

\begin{figure}
\narrowtext
\epsfxsize=2.4in\epsfysize=2.4in
\hskip 0.3in\epsfbox{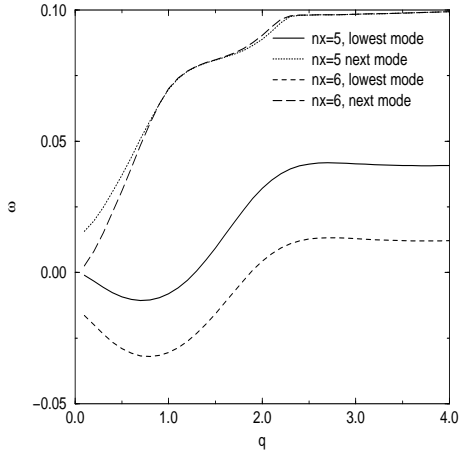}
\vskip 0.15in
\caption{Lowest and next lowest spin-flip modes for $\nu=\twof$ 
at $\lam=0.0$ for different values of $n_{max}$. Note that the energy
of the lowest mode is negative for a substantial range of $q$, and that there
is a roton-like minimum.
\label{fig17}}
\end{figure}

\begin{figure}
\narrowtext
\epsfxsize=2.4in\epsfysize=2.4in
\hskip 0.3in\epsfbox{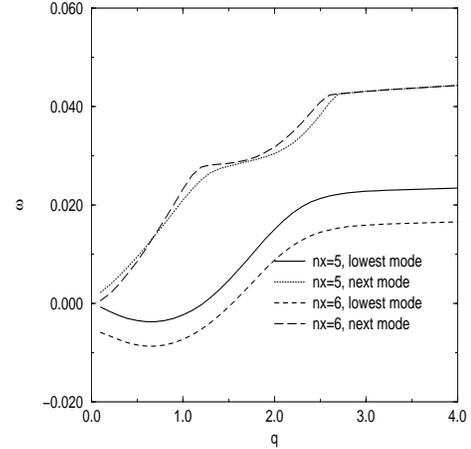}
\vskip 0.15in
\caption{Lowest and next lowest spin-flip modes for $\nu=\twof$ 
at $\lam=1.0$.  All the features observed for $\lam=0.0$ are present.
\label{fig18}}
\end{figure}

Next we present the spin-flip excitations for $\twof$, and here there
is a new feature. While there is always a gapless quadratically
dispersing mode, it is not always the lowest mode. For $n_{max}=5$ the
gapless mode has negative energy for small $q$, while for $n_{max}=6$,
the gapless mode has positive energy, but the negative energy mode
persists, and becomes gapped.  The lowest mode at large $q$ is one
which takes an up-spin particle from the $n=1$ CF-LL and replaces it
as a down-spin particle in the $n=0$ CF-LL. At large $q$ the energy of
this excitation is positive, making the spin-reversed gap
positive. However, as shown in Figs. \ref{fig17} and \ref{fig18},
there is always a negative energy mode at small $q$. In the absence of
a Zeeman coupling  this implies an instability of the polarized ground
state. It is known from previous work\cite{park2} comparing ground
state energies that the spin-singlet $\twof$ state is lower in energy
than the spin-polarized state for small $E_Z$. The zero-temperature
transition between the states is expected to be first order, but
nonetheless, one can look for signatures of the instability in the
polarized state.  

In fact, from Fig. \ref{fig18} we can infer that at
$\lam=1.0$ (presumably close to the physical range of sample
thickness), the transition to the polarized state should take place
when a uniform shift of the ME dispersion upwards by $E_Z$ makes the
dispersion always positive, {\it i.e.}, for  ${E_Z\over
e^2/\varepsilon l}\ge0.01$. Experimentally\cite{du-spin}, it is found
that $E_Z\approx 0.015 e^2/\varepsilon l$.

In the presence of a Zeeman energy $E_z$ capable of stabilizing a
polarized ground state, our result implies that the lowest spin-flip
mode will have a roton-like minimum at finite $q$, and that the
smallest gap to excitations is lower than $E_z$. This in turn would
indicate a much faster reduction of the magnetization with temperature
than in the case of $\third$, and a finite-temperature spin-structure
factor peaked at nonzero momentum. Perhaps the most significant
consequence of this result is that one can no longer think of the
polarized $\twof$ state as a simple quantum ferromagnet.

\bigskip
\section{Conclusions}
\bigskip

I have used the dynamical theory of Composite Fermions developed by
Shankar and myself\cite{us1,us2} to compute the magnetoexciton
dispersions for a few gapped fractions.  The TDHF method of
MacDonald\cite{macd1,macd2} was used to do these computations. Let me
reiterate that this method is the minimal one in the present case
since there is only one energy scale. My primary motivation is to see
how far our theory can take us, and what its limitations are. To this
end, the constraint was imposed in its crudest form, by cutting off
the CF-LLs at $n_{max}$. Some such cutoff is certainly necessary on
physical grounds, as explained in section II.

Our results for spin-polarized ME dispersions are in qualitative
agreement with exact diagonalizations\cite{hal-rez,exact-ex}, and
other numerical methods\cite{GMP,jain-ex}, for the pure Coulomb
interaction ($\lam=0.0$). This is the worst case for our theory, since
our expression for the charge density is valid at long wavelengths,
and the Coulomb interaction has a significant short-distance part. As
a consequence, there are significant numerical discrepancies for
$\lam=0.0$.

On the other hand, as the sample thickness increases, the interaction
becomes softer at high momenta, and our theory should become more
accurate. To model this, we considered a particular interaction with a
parameter $\Lambda=\lam l$ related to sample thickness. I have
presented a number of results at larger $\lam$. In particular, the
cutoff dependence becomes negligible at large $\lam$, increasing our
confidence in the results. Generally, we find that while the structure
of minima found in the ME dispersions for the pure Coulomb case
($\lam=0.0$) persists at larger $\lam$, the minima become shallower
(and even disappear for $\third$ around $\lam=3.0$), while the entire
ME dispersion moves down in energy. We also find that the fraction of
the spectral weight in the lowest mode decreases with sample
thickness. Finally, we find that the $\twof$ incompressible state
should become unstable somewhere around $\lam_{cr}=2.65$, while the
$\threes$ incompressible state should disappear near
$\lam_{cr}=2.4$. The instabilities are at finite wavevector, and one
may speculate that the transition might be to a crystal
phase\cite{wc-instability}. These are estimates for a clean system,
and would be revised downwards for systems with
disorder. Unfortunately, results from exact diagonalizations or other
essentially exact methods are not currently available to test our
results for thick samples.

Turning to the spin-flip MEs, we find good agreement between our
results and exact diagonalizations\cite{nakajima} for the case
$\nu=\third$, even in the worst case $\lam=0.0$. We find that for
$\third$, the spin-wave mode is the lowest one, and is well separated
from the other modes for small $q$, justifing its treatment as a
quantum ferromagnet\cite{q-fm-th}. However, for $\nu=\twof$ we find
that the lowest mode is {\it not} the spin-wave, but a spin-flip
excitation that creates a hole in the $n=1$ CF-LL, and puts a minority
spin particle in the $n=0$ CF-LL. This implies that a quantum
ferromagnet treatment\cite{q-fm-th} that ignores the CFs would be
inadequate in this case. Furthermore, the lowest excitation has negative
energy (in the absence of a Zeeman interaction) for a substantial range
of $q$, with a minimum at finite wavevector. These predictions should
be detectable in the spin structure factor, and in the temperature
dependence of the magnetization of the polarized $\nu=\twof$ state. 

The most significant deficiency in our treatment is the way we handle
the constraints. Constraints are known to be
crucial\cite{hlr,dh,read3,comment,stern} for compressible systems such
as $\nu=\half$. Ideally, one would like to employ a conserving
approximation\cite{conserving} in which the constraint would be
satisfied, such as the one carried out by Read for the compressible
states\cite{read3}. An LLL formulation for all the principal fractions
valid at all $q$, which is algebraically consistent, and which agrees
with our approach at small $q$, has recently been postulated by
Shankar\cite{aftermath}. This formulation is amenable to a conserving
approximation, and would be the way to settle matters of
principle. However, we presented several lines of reasoning to suggest
that in using our $\rhob$ we are already building in many nontrivial
renormalizations at tree level, and that the effect of constraints
would be minimal. The formulation of Shankar\cite{aftermath} shares
this philosophy, and provides a partial justification of our
truncation of $\rhob$ (in the smallness of the higher order terms of
$\rhob_{pref}$\cite{aftermath}).

In conclusion, our dynamical theory of Composite Fermions gives a
satisfactory approximate account of the low-energy excitations around
the gapped FQH states, the magnetoexciton modes. In addition to
reproducing previous results we find some surprising
predictions. There are many directions in which one can take this
approach. Perhaps the most important are the treatment of
disorder\cite{disorder}, and an understanding of the crystal
phase\cite{wc-review}, which I hope to return to in future work.

\section{Acknowledgements}
I would like to thank J.K.Jain, R.Shankar, S.H.Simon, and S.L.Sondhi 
for helpful conversations.


\begin{thebibliography}{100}
\bibitem{fqhe-ex} D.Tsui, H.Stromer and A.Gossard, Phys. Rev. Lett.
{\bf 48},
1599, (1982).
\bibitem{laugh} R.B.Laughlin,  Phys. Rev. Lett. {\bf 50},
1395, (1983).
\bibitem{statistics} B.I.Halperin, \prl {\bf 52}, 1583 (1984); 
D.P.Arovas, J.R.Schreiffer, and F.Wilczek, \prl {\bf 53}, 722 (1984).
\bibitem{jain-cf} J.K.Jain, Phys. Rev. Lett.  {\bf 63}, 199, (1989);
Phys. Rev.
{\bf B 41}, 7653 (1990); Science {\bf 266}, 1199 (1994).
\bibitem{jain-cf-review} J.K.Jain and R.K.Kamilla, in Chapter 1 of   
``Composite
Fermions'', Olle Heinonen, Editor (World Scietific, Teaneck, NJ, 1998).
\bibitem{stormer-cf} For a review, see H.L.Stormer and D.C.Tsui in Chapter 10 of ``Perspectives in Quantum Hall Effects'', S.Das Sarma and A.Pinczuk
editors (Wiley, New York, 1997).
\bibitem{cs-trans} J.M.Leinaas and J.Myrheim, 
Nuovo Cimento {\bf 37B}, 1 (1977). 
\bibitem{gcs} S.M.Girvin, in Chapter 9 of ``The Quantum Hall
Effect'',
Edited by R.E.Prange and S.M Girvin, Springer-Verlag, 1990.
\bibitem{gmcs} S.M.Grivin and A.H.MacDonald,
Phys. Rev. Lett. {\bf 58}, 1252 (1987).
\bibitem{zhk}S.-C.Zhang, H.Hansson and S.A.Kivelson, Phys. Rev. Lett.
{\bf 62}, 82, (1989); D.-H.Lee and S.-C.Zhang, Phys. Rev. Lett. {\bf
66}, 1220 (1991); S.-C.Zhang, Int. J. Mod. Phys., {\bf B6}, 25
(1992).
\bibitem{read1} N.Read, Phys. Rev. Lett., {\bf 62}, 86 (1989).
\bibitem{lopez} A.Lopez and E.Fradkin, Phys. Rev. B {\bf  44}, 5246
(1991), {\em ibid} {\bf 47}, 7080, (1993), Phys. Rev. Lett.  {\bf 69},
2126 (1992); For the latest review of the fermionic CS approach see
A.Lopez and E.Fradkin in Chapter 3 of ``Composite Fermions'', Olle
Heinonen, Editor, (World Scietific, Teaneck, NJ, 1998).
\bibitem{kalmeyer} V.Kalmeyer and S.-C.Zhang, Phys. Rev. B {\bf 46},
9889 (1992).
\bibitem{hlr} B.I.Halperin, P.A.Lee and N.Read, Phys. Rev. B {\bf
47}, 7312 (1993).
\bibitem{us1} R.Shankar and G.Murthy, Phys. Rev. Lett. {\bf 79},
4437, (1997).
\bibitem{us2} G.Murthy and R.Shankar, in Chapter 4 of ``Composite
Fermions'', Olle Heinonen, Editor, (World Scietific, Teaneck, NJ,
1998).
\bibitem{bohm-pines} D.Bohm and D.Pines, Phys. Rev. {\bf 92}, 609,
(1953).
\bibitem{ssh} S.H.Simon, A.Stern, and B.I.Halperin, Phys. Rev. B {\bf
54},
R11114 (1996).
\bibitem{simon-f1} S.H.Simon and B.I.Halperin, Phys. Rev. B {\bf 48},
17368, (1993).
\bibitem{read2} N.Read, Semi. Sci. Tech. {\bf 9}, 1859 (1994);
Surf. Sci., {\bf 361/362}, 7 (1996).
\bibitem{kohn} W.Kohn, Phys. Rev {\bf 123}, 1242 (1961).
\bibitem{GMP} S.M.Girvin, A.H.MacDonald and P.Platzman, Phys. Rev.
B
{\bf 33},
2481, (1986).
\bibitem{dh} D.H.Lee, \prl {\bf 80}, 4745 (1998). \label{lee}
\bibitem{pasquier} V.Pasquier and F.D.M.Haldane, Nucl. Phys. {\bf B516}, 719 (1998). 
\bibitem{read3} N.Read, \prb {\bf 58}, 16262 (1998).
\bibitem{thick1} F.F.Fang and W.E.Howard, Phys. Rev. Lett. {\bf 16},
797 (1966); T.Ando, A.B.Fowler, and F.Stern, Rev. Mod. Phys. {\bf 54},
437 (1982).
\bibitem{hal-rez} F.D.M.Haldane and E.H.Rezayi, Phys. Rev. Lett. {\bf
54}, 237
(1985).
\bibitem{thick2} F.C.Zhang and S.Das Sarma, Phys. Rev. B {\bf 33},
2903
(1986);
D.Yoshioka, J. Phys. Soc. Jpn. {\bf 55}, 885 (1986); T.Chakraborty,
P.Pietlainen, and F.C.Zhang, Phys. Rev. Lett, {\bf 57}, 130 (1986);
R.Morf and B.I.Halperin, Z. Phys. B {\bf  68}, 391 (1987); R.Price
and
S.Das Sarma, Phys. Rev. B {\bf  54}, 8033 (1996).
\bibitem{bonesteel} V.Melik-Alaverdian and N.E.Bonesteel, Phys. Rev.
B
{\bf 52},
R17032 (1995); V.Melik-Alaverdian, N.E.Bonesteel, and G.Ortiz, Phys.
Rev.
Lett., {\bf 79}, 5286 (1997).
\bibitem{jain-th} K.Park and J.K.Jain, \prl {\bf 81}, 4200 (1998).
\bibitem{single-part} G.Murthy and R.Shankar, cond-mat 9806380, 
to appear in \prb. 
\bibitem{scaling} G.Murthy, K.Park, R.Shankar, and J.K.Jain, \prb {\bf 58}, 13263 (1998).
\bibitem{me-history} K.W.Chiu and J.J.Quinn, Phys. Rev. B {\bf 9}, 4724
(1974); N.Horing and M.Yildiz, Ann. Phys. NY {\bf 97}, 216 (1976);
T.Theis, Surf. Sci. {\bf 98}, 515 (1980); P.Hawrylak and J.J.Quinn,
\prb {\bf 31}, 6592 (1985); Yu.A.Bychkov, S.V.Iordanskii, and
G.M.Eliashberg, JETP Lett. {\bf 33}, 152 (1981).
\bibitem{kallin} C.Kallin and B.I.Halperin Phys. Rev B 30, 5655, (1984).
\bibitem{macd1} A.H.MacDonald, \jpc {\bf 18}, 1003 (1985).
\bibitem{macd2} H.C.A.Oji and A.H.MacDonald, \prb {\bf 33}, 3810 (1986); 
I.K.Marmorkos and S.Das Sarma, \prb {\bf 45}, 13396 (1992); J.P.Longo
and C.Kallin, \prb {\bf 47}, 4429 (1993).
\bibitem{exact-ex} N.d'Ambrumenil and R.Morf, \prb {\bf 40}, 6108
(1989); P.M.Platzman and S.He, \prb {\bf 49}, 13674 (1994); S.He,
S.H.Simon, and B.I.Halperin, \prb {\bf 50}, 1823 (1994); S.He and
P.M.Platzman, Surf. Sci {\bf 361/362}, 87 (1996); S.He, Computers in
Phys. {\bf 11}, 194 (1997).
\bibitem{jain-ex} X.G.Wu and J.K.Jain, \prb {\bf 51}, 1752 (1995);
R.K.Kamilla, X.G.Wu, and J.K.Jain, \prl {\bf 76}, 1332 (1996); \prb
{\bf 54}, 4873 (1996).
\bibitem{rpa-ex} S.H.Simon and B.I.Halperin, \prb {\bf 48}, 17368 (1993); S.H.Simon and B.I.Halperin, 
Phys. Rev. B {\bf 50}, 1807 (1994); X.C.Xie, \prb {\bf 49}, 16833 (1994).
\bibitem{shivaji-skyrmion} S.L.Sondhi, A.Karlhede, S.A.Kivelson, and 
E.H.Rezayi, \prb {\bf 47}, 16419 (1993).
\bibitem{q-fm-th} N.Read and S.Sachdev, \prl {\bf 75}, 3509 (1995); 
M.Kasner and A.H.MacDonald, \prl {\bf 76}, 3204 (1996); C.Timm,
S.M.Girvin, P.Henlius, and A.W.Sandvik, \prb {\bf 58}, 1464 (1998);
M.Kasner, J.J.Palacios, and A.H.MacDonald, cond-mat 9808186.
\bibitem{q-fm-ex} R.Tycko, S.E.Barrett, G.Dabbagh, L.N.Pfeiffer, 
and K.W.West, Science {\bf 268}, 1460 (1995); S.E.Barrett, G.Dabbagh,
L.N.Pfeiffer, K.W.West, and R.Tycko, \prl {\bf 74}, 5112 (1995);
M.J.Manfra, E.H.Aifer, B.B.Goldberg, D.A.Broido, L.N.Pfeiffer, and
K.W.West, \prb {\bf 54}, 17327 (1996).
\bibitem{nakajima} T.Nakajima and H.Aoki, \prl {\bf 73}, 3568 (1994).
\bibitem{mandal} S.S.Mandal, \prb {\bf 56}, 7525 (1997). 
\bibitem{wc-instability} P.K.Lam and S.M.Girvin, \prb {\bf 30}, 473 (1984); 
D.Levesque, J.J.Weiss, and A.H.MacDonald, \prb {\bf 30}, 1056 (1984). 
\bibitem{jain-wc} R.K.Kamilla and J.K.Jain, \prb {\bf 55}, R13417 (1997).
\bibitem{fetter} A.L.Fetter, C.B.Hanna, and R.B.Laughlin, Phys.
Rev. B {\bf 39}, 9679 (1989); {\em ibid} {\bf 43}, 309 (1991); Q.Dai,
J.L.Levy, A.L.Fetter, C.B.Hanna, and R.B.Laughlin, Phys. Rev. B {\bf
46}, 5642 (1992).
\bibitem{lerner} I.V.Lerner and Yu.E.Lozovik, Sov. Phys. JETP {\bf
57}, 588 (1980).
\bibitem{aftermath} R.Shankar, cond-mat 9903064.  
\bibitem{haldane-book} F.D.M.Haldane, in Chapter 8 of 
``The Quantum Hall Effect'', R.E.Prange and S.M.Girvin, Editors
(Springer-Verlag, New York, 1990).
\bibitem{dev} G.Dev and J.K.Jain, Phys. Rev. Lett. {\bf 69}, 2843
(1992).
\bibitem{comment} B.I.Halperin and A.Stern, \prl {\bf 80}, 5457 (1998); G.Murthy and R.Shankar, {\it ibid} 5458 (1998).
\bibitem{stern} A.Stern, B.I.Halperin, F.von Oppen, and S.H.Simon, 
cond-mat 9812135.
\bibitem{conserving} G.Baym and L.P.Kadanoff, 
Phys. Rev. {\bf 124}, 287 (1961). 
\bibitem{operator-rpa} G.Rickayzen, Phys. Rev. {\bf 115}, 795 (1957). 
\bibitem{morf-comment} R.H.Morf, cond-mat 9812181. Morf raises the 
possibility that the values of $\lam$ used in ref.\cite{jain-th} to
compare to the experimental numbers or ref.\cite{du-numbers}, are
incorrect due to errors in ref.\cite{ortalano}. He suggests that the
physical values of $\lam$ are between $1$ and   $1.5$.
\bibitem{du-numbers} R.R.Du, H.L.St\"ormer, D.C.Tsui, L.N.Pfeiffer,
and K.W.West, \prl {\bf 70}, 2944 (1993).
\bibitem{ortalano} M.W.Ortalano, S.He, and S.Das Sarma, \prb {\bf
55},
7702 (1997).

\bibitem{moon} K.Moon, H.Mori, K.Yang, S.M.Girvin, A.H.MacDonald, L.Zheng, D.Yoshioka, and S.-C.Zhang, \prb {\bf 51}, 5138 (1995). 
\bibitem{hal-spin} B.I.Halperin, Helv. Phys. Acta {\bf 556}, 75 (1983). 
\bibitem{lopez2} A.Lopez and E.Fradkin, \prb {\bf 51}, 4347 (1995).
\bibitem{double-layer-th} K.Yang, K.Moon, L.Zheng, A.H.MacDonald, 
S.M.Girvin, D.Yoshioka, and S.-C.Zhang, \prl {\bf 72}, 732 (1994); For
a review, see S.M.Girvin in Chapter 5 of ``Perspectives in Quantum
Hall Effects'', S.Das Sarma and A.Pinczuk editors (Wiley, New York,
1997).
\bibitem{double-layer-ex} For a
 review of the experiments, see J.P.Eisenstein in Chapter 2 of
``Perspectives in Quantum Hall Effects'', S.Das Sarma and A.Pinczuk
editors (Wiley, New York, 1997).
\bibitem{park2} K.Park and J.K.Jain, \prl {\bf 80}, 4237 (1998).  
\bibitem{du-spin} R.R.Du, A.S.Yeh, H.L.Stormer, D.C.Tsui, L.N.Pfeiffer, 
and K.W.West, \prl {\bf 75}, 3926 (1995). 
\bibitem{disorder} R.B.Laughlin, M.L.Cohen, J.M.Kosterlitz, H.Levine, 
S.B.Libby and A.M.M.Pruisken, \prb {\bf 32}, 1311 (1985);
A.H.MacDonald, K.L.Liu, S.M.Girvin, and P.M.Platzman, \prb {\bf 33},
4014 (1986).
\bibitem{wc-review} For a review see H.A.Fertig in Chapter 3 of 
``Perspectives in Quantum Hall Effects'', S.Das Sarma and A.Pinczuk
editors (Wiley, New York, 1997).





\end{thebibliography}
\end{document}